\documentclass{iopart}

\usepackage{graphicx,color}

\usepackage{placeins}

\newcommand{\sgn}{\ensuremath{\mbox{sgn}}}
\newcommand{\artanh}{\ensuremath{\mbox{artanh}}}

\graphicspath{{../fig/}}

\begin{document}
\title{Nonlocal electrodynamics of normal and superconducting films}

\author{J~I~Vestg{\aa}rden$^1$, P~Mikheenko$^1$, Y~M~Galperin$^{1~2}$, T~H~Johansen$^{1~3}$ }
\address{$^1$ Department of Physics, University of Oslo, PO Box 1048
Blindern, 0316 Oslo Norway}
\address{$^2$ Ioffe Physical Technical Institute, 26 Polytekhnicheskaya, 
St Petersburg 194021, Russian Federation}
\address{$^3$  Institute for Superconducting and Electronic Materials,
University of Wollongong, Northfields Avenue, Wollongong, NSW 2522,  Australia}
\ead{j.i.vestgarden@fys.uio.no}

\begin{abstract}
Electrically conducting films in a time-varying transverse applied
magnetic field are considered. Their behavior is strongly influenced by the
self-field of the induced currents, making the electrodynamics nonlocal,
and consequently difficult to analyze both numerically and analytically.
We present a formalism which allows many 
phenomena related to superconducting and Ohmic films to be modelled and analyzed.
The formalism is based on the Maxwell equations, and a material
current-voltage characteristics, linear for normal metals, and nonlinear
for superconductors, plus a careful account of the boundary conditions.
For Ohmic films, we consider the response to 
a delta function source-field turned on instantly.  As one of
few problems in nonlocal electrodynamics, this has an analytical
solution, which we obtain, in both Fourier and real space.  
Next, the dynamical behavior of a square superconductor film during ramping up of
the field, and subsequently returning to zero, is treated numerically. 
Then, this remanent state is used as initial condition for triggering
thermomagnetic avalanches. The avalanches tend to invade the central part
where the density of trapped flux is largest, forming dendritic patterns
in excellent agreement with magneto-optical images. Detailed profiles of
current and flux density are presented and discussed.  Finally, the formalism
is extended to multiply connected samples, and 
numerical results for a patterned superconducting film, a ring with a square lattice
of antidots, are presented and discussed.
\end{abstract}


\maketitle


\section{Introduction}
\label{sec:introduction}
The flux dynamics in electrically conducting films experiencing a
time-varying transverse applied magnetic field is governed by the
Maxwell equations, with the material characteristics supplied as an
additional relation between the electric field and current density.
The systems to be addressed in this work range from superconductors to Ohmic
materials.  To solve these equations it is necessary to determine the
currents induced in the film as the magnetic field varies. This is a
nontrivial task since one must also account for the significant
self-field of the induced currents, which makes the final relations
nonlocal \cite{brandt95}.

The electromagnetic behaviour of type-II superconductors is often well
described by Bean's critical-state model \cite{bean64}. For bulk
samples initially zero-field-cooled below the transition temperature,
$T_{\rm{c}}$, and then exposed to an increasing applied magnetic
field, $H_{\rm{a}}$, the model tells that the material sets up
lossfree shielding currents of critical density, $j_{\rm{c}}$. This
current flows in the same macroscopic regions as where the magnetic
flux is allowed to penetrate, while the inner unpenetrated part
remains free of currents.  In films, on the other hand, the
electromagnetic non-locality implies that induced currents flow in 
the entire sample \cite{norris69,brandt93}.  Thus, the film behaviour is qualitatively
different from that of bulks, and magneto-optical imaging (MOI) of
thin superconductors has revealed strong piling up of the magnetic
field around the sample edges, where values far above $H_{\rm{a}}$ are
reached \cite{jooss02}.  At internal boundaries, such as the inner
edge of a planar ring, the field can, 
due to the nonlocal electrodynamics, be in opposite direction
of the applied field
\cite{brandt97,pannetier01}.  Strongly modified behaviour is found
also in films patterned with regular arrays of small holes (antidots),
which tend to guide the flux into the superconductor
\cite{pannetier03,gheorghe06,tamegai10, vestgarden12}.

The response of Ohmic films exposed to varying transverse magnetic
fields is also described by nonlocal electrodynamics, but here the
material responds linearly.  Numerical solutions for strip and disk
geometries have shown that the combination of non-locality and
dissipation causes a rapid penetration of a suddenly applied magnetic
field \cite{brandt93-prl,brandt94-disk}.  Different from
superconductors, even regions deep inside an Ohmic film are quickly
penetrated by the magnetic field.

A phenomenon that involves both the critical-state and Ohmic
properties is the occurance of flux avalanches or flux jumps. These
are commonly observed in type-II superconductors at low temperatures,
and are caused by a thermomagnetic instability which drives the
superconductor from the critical-state to a high resistivity state
\cite{mints81}.  The instability is triggered, e.g., by a small
temperature fluctuation which reduces the flux pinning locally, and
some quantized flux lines, or vortices, will start moving. This
creates local heat dissipation and the temperature will increase even
further, thus forming a positive feedback loop.  The result can be an
exponential growth in the temperature and a large-scale runaway of
magnetic flux.  In superconducting films the thermomagnetic
instability is seen by MOI to manifest as abrupt avalanches of
magnetic flux which form complex branching filamentary structures,
socalled dendritic avalanches
\cite{duran95,wimbush04,choi05,rudnev05,denisov06,yurchenko07,treiber11}.

These avalanches can be modeled using the equations describing
nonlocal and nonlinear electrodynamics coupled with an equation for
the production and propagation of heat \cite{aranson01}. Linearization
of the equations has been highly succesful in parametrizing the
conditions for onset of the instability, confirming that the nonlocal
electrodynamics makes a significant difference between bulk
\cite{rakhmanov04} and film geometries
\cite{mints96,denisov05,denisov06,aranson05,vestgarden13-metal}. Numerical
simulations of the full time evolution of the avalanches have produced
dendritic flux patterns in excellent agreement with the experimental
MOI results \cite{aranson05,vestgarden11}.  The propagation of these
avalanches is extremely fast -- velocities up tp 180 km/s have been
measured \cite{bolz03}, and the process is driven by adiabatic heating
\cite{vestgarden12-sr}.  During an avalanche the local temperature is
expected to rise above $T_{\rm{c}}$, thus bringing for a very short
time interval the superconductor to the normal conducting state. In
such cases, the process is governed by the interplay between
supercurrents and Ohmic normal-state currents.

\begin{figure}[t]
  \centering  
  \includegraphics[width=10cm]{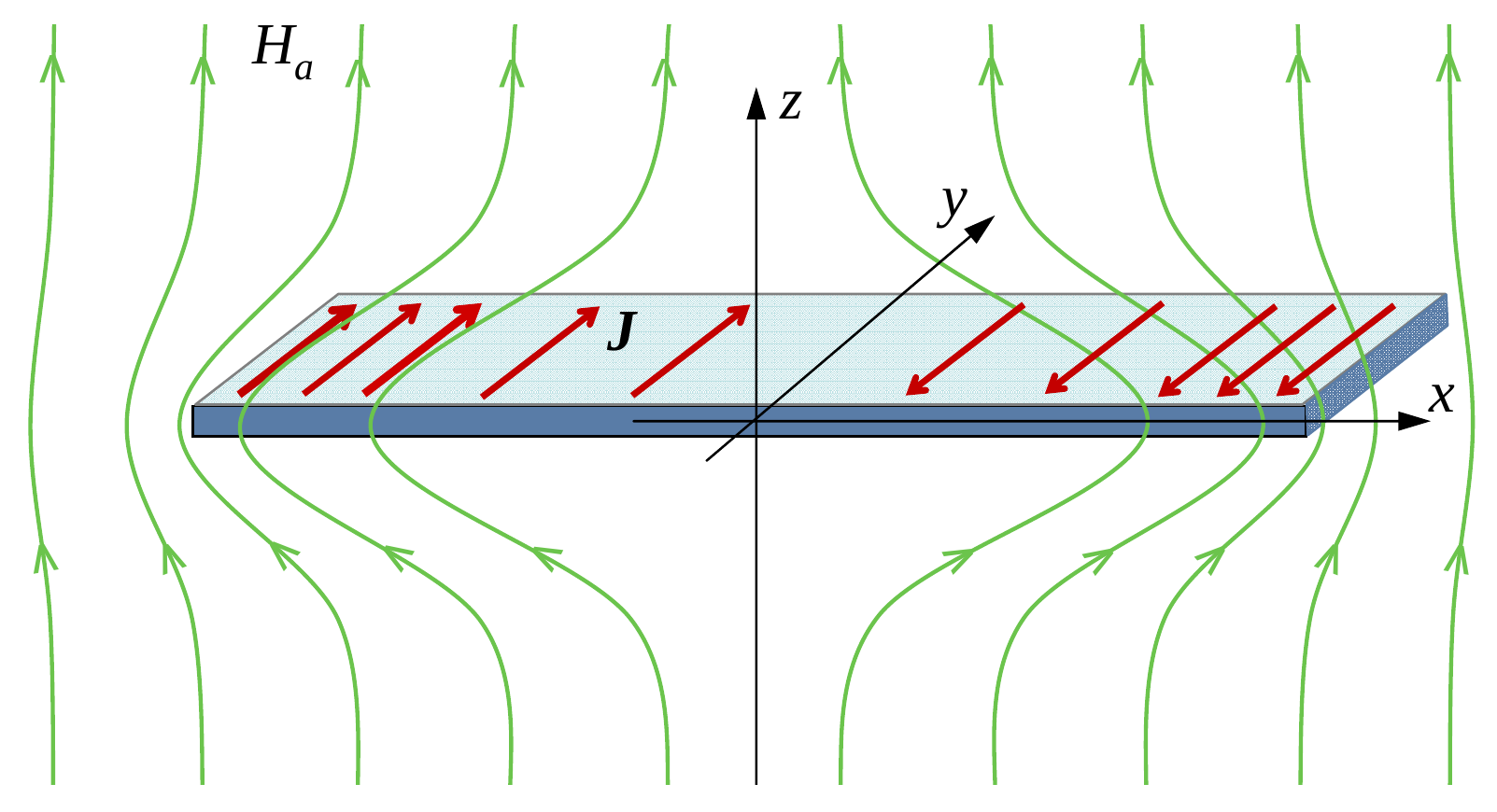} \\
  \caption{
    \label{fig:fig1} 
    An electrically conducting film in time-varying transverse applied magnetic field $H_{\rm{a}}(t)$.
    Due to the induced current, $\bi{J}$, the sample partly expels the transverse field component $H_{\rm{z}}$.
  }
\end{figure}

In this work, we consider the electrodynamics of normal and superconducting 
films in transverse applied magnetic field; see figure~\ref{fig:fig1}. 
The basic idea is that a wide range of physical problems in this geometry 
can be described by the same formalism based on the Maxwell equations, 
only by supplying a relation between electric field $E$ and sheet 
current $J$ to characterize the material. Thus, we will 
describe the formalism in detail, with particular focus on enforcement of 
the boundary conditions. Having described the formalism,
we apply the solution method to various physical problems. 
As a first example, we calculate the remanent flux in 
a superconducting square, after the applied magnetic field has been increased 
to reach full penetration, and then decreased back to zero.
Due to the nonlocal electrodynamics, the field and current distributions 
in the square are highly nontrivial. From the remanent state, we consider 
the evolution of dendritic flux avalanches, which means that we must model
the propagation of heat in the system, in addition to the electrodynamics.
Our numerical solution is compared with a magneto-optical imaging experiment 
which maps the magnetic flux distribution in a NbN superconductor in descending magnetic field.
As a separate problem, we considers the response of an infinite Ohmic sheet 
to a delta function source field. This problem is analytically solvable and 
the solution sheds light on the dynamics of Ohmic films, as well as the dynamics
of dendritic flux avalanches, which is driven by a normal domain invading 
a superconducting phase.
Finally, we consider a superconducting ring patterned with a regular array of 
antidots. This system is interesting due to the conflicting symmetries of 
the sample and the antidot array, but rather difficult to handle numerically due 
to the complicated sample layout. In total, all these problems demonstrate 
that our formalism is powerful and flexible, as it can be applied to 
a wide range of physical problems in the thin-film transverse geometry.

This paper is organized as follows.
Section \ref{sec:model} describes the transverse geometry.
Section \ref{sec:creep} finds the remanent flux distribution in a square superconductor,
and calculates also the analytical solution for the field and currents in a 
normal metal film subjected to delta function source field.
Section \ref{sec:dendrite} considers dendritic flux avalanches in the remanent state,
both numerically and by a magneto-optical imaging experiment.
Section \ref{sec:antidot} considers the dynamics of a superconducting ring patterned with
antidots.
Finally, section \ref{sec:conclusion} provides the conclusion.

\section{The transverse geometry}
\label{sec:model}

\subsection{Connecting magnetic field and current distributions}   
A key element in solving magnetic flux dynamics problems in films
placed in a transversely applied field, $H_{\rm{a}}$, both Ohmic and
superconducting ones, is the relation between the distributions of
electrical current and transverse magnetic field $H_{\rm{z}}(x,y)$ over the
$(x,y)$-plane defined by the film.  To establish the formalism 
used in this work, we assume the film thickness to be much smaller than any length
characterizing the patterns of flux and currents.  The current density
in the film can then be expressed as
\begin{equation}
  \bi j = \bi J(x,y) \delta(z) ,
\end{equation}
where $\bi J$ is the sheet current.
It is convenient to introduce the local magnetization $g=g(x,y)$ as
\begin{equation}
  \bi J = \nabla\times  \hat z g \,  .
\end{equation}
where $\hat z$ is the unit vector transverse to the sample plane. 
The total magnetic moment of the film can then be expressed as
\begin{equation}
  m\hat z=\frac{1}{2}\int\bi r\times \bi j(\bi r) \rmd^3r
  =\hat z\int g \,\rmd x \rmd y .
\end{equation}
Neglecting the displacement field, the Amp\`ere law becomes 
\begin{equation}
  \nabla\times \bi H 
  = \left(\nabla \times \hat z g\right)\delta(z) ,
\end{equation}
and Fourier transforms along the Cartesian axes give 
\begin{eqnarray*}
  & \rmi k_{\rm{y}}H^{[3]}_z-\rmi k_zH^{[3]}_{\rm{y}}=\rmi k_{\rm{y}}g^{[2]}, \\ -&
  \rmi k_{\rm{x}}H^{[3]}_z+\rmi k_zH^{[3]}_{\rm{x}}=-\rmi k_{\rm{x}}g^{[2]}, \\ &
  \rmi k_{\rm{x}}H^{[3]}_{\rm{y}}-\rmi k_{\rm{y}}H^{[3]}_{\rm{x}}=0.
\end{eqnarray*}
Here $\bi H^{[3]} = \bi H^{[3]}(k_{\rm{x}},k_{\rm{y}},k_z)$ is the
three-dimensional Fourier transform of $\bi H$ 
and $g^{[2]} = g^{[2]}(k_{\rm{x}},k_{\rm{y}})$.  
Conservation of magnetic flux, $\nabla \cdot \bi H=0$,
yields 
\begin{equation*}
  \rmi k_{\rm{x}}H^{[3]}_{\rm{x}}+\rmi k_{\rm{y}}H^{[3]}_{\rm{y}}+\rmi k_zH^{[3]}_z=0 
  ,
\end{equation*}
so that 
\begin{equation*}
  H^{[3]}_{\rm{x}} = \frac{\rmi k_{\rm{x}}\rmi k_z}{k^2}H^{[3]}_z ,
\end{equation*}
where $k=\sqrt{k_{\rm{x}}^2+k_{\rm{y}}^2}$. Thus, $H_{\rm{x}}$ is nonzero, and the same
holds for $H_{\rm{y}}$, which is a general feature of films in the transverse 
geometry. Isolating $H_{\rm{z}}^{[3]}$ gives
\begin{equation*}
  H^{[3]}_{\rm{z}} = \frac{k^2}{k_{\rm{z}}^2+k^2}g^{[2]} ,
\end{equation*}
and inverse Fourier transform in $z$ direction results in the final expression
\begin{equation}
  H^{[2]}_{\rm{z}}= \frac{k}{2}\rme^{-k|z|}g^{[2]} ,
  \label{ampere1}
\end{equation}
where $H_{\rm{z}}^{[2]}=H^{[2]}(k_{\rm{x}},k_{\rm{y}},z)$. For inversion,  e.g., of magneto-optical
images \cite{johansen96, baziljevich96-2, gaevski99} one often uses a finite $z$ 
to account for a small gap between the sample and the field sensing, i.e., 
Faraday rotating layer. However, for the flux dynamics calculations 
in this work we only consider the expressions at $z=0$.

The  $H_{\rm{z}} - g$ relation will henceforth be denoted  
as the Biot-Savart law, and it can be written as \cite{roth89}
\begin{equation}  
  \label{def-Q}
  H_{\rm{z}}(x,y)=\hat{Q}\left[g(x,y)\right] \equiv
  \mathcal{F}^{-1}\left[\frac{k}{2}\mathcal{F}\left[g(x,y)\right] \right]
  ,
\end{equation}
where $\mathcal F$ and $\mathcal F^{-1}$  is forward and inverse
Fourier transform, respectively.  The inverse relation is equally
simple,
\begin{equation}
  \label{def-invQ}
  g(x,y)= \hat{Q}^{-1}\left[H_{\rm{z}}(x,y)\right] \equiv
  \mathcal{F}^{-1}\left[\frac{2}{k}\mathcal{F}\left[H_{\rm{z}}(x,y)\right] \right] 
  .
\end{equation}
The above equations are exact on an infinite sheet. For
 films having a finite area, they are good 
approximations for short wavelengths \cite{brandt95}.

\subsection{Iteration scheme}
\label{subsec:scheme}
Consider a planar conducting film surrounded by vacuum, and with $H_{\rm{z}}$
known inside the sample area defined by its boundary. Given the task
to determine the local magnetization, $g$, the intuitive approach is
to use (\ref{def-invQ}). However, this fails to give correct
result unless $H_{\rm{z}}$ is known over the entire plane.  An approach
allowing $g$ to be found correctly was invented by
Brandt \cite{brandt95,brandt01}, and is based on a matrix inversion
scheme. The approach proved to work very well for simple geometries
which can be represented by a fairly small number of discrete grid
points.  Later, the numerical performance of the matrix inversion was
improved by using a congruent gradient method
\cite{wijngaarden98,loerincz04}.

An alternative approach is to try and extrapolate $H_{\rm{z}}(x,y)$ to 
the outside area, and then apply (\ref{def-invQ}). For an infinitely long
strip, this can be done by symmetry considerations, as shown by
Aranson et al. \cite{aranson05}.  In this work we consider far more
general geometries, and will calculate $H_{\rm{z}}$ making use of the fact
that outside the sample one has $g=0$. Our scheme is iterative,
and as will be demonstrated, computationally efficient
\cite{vestgarden11}.

To describe our approach, it is convenient to define a function
representing the projection on the sample,
\begin{equation}
  \label{defS}
  S(x,y) = 
  \left\{
  \begin{array}{ll}
    1,&  \mbox{ inside the sample boundary}, \\
    0,& \mbox{ outside the sample boundary}.     
  \end{array}
  \right.
\end{equation}
The corresponding projection on the outside region is $1-S(x,y)$.
For brevity, the argument $(x,y)$ is omitted in the next expressions. 
The iterations start by defining a trial function $H_{\rm{z}}^{(i)}$, which has 
correct values inside the sample, i.e., $  SH_{\rm{z}}^{(i)} = SH_{\rm{z}},$
and given by an initial guess for the field outside, $(1-S)H_{\rm{z}}^{(i)}  $.
The quantities to be determined by iterations are $(1-S)H_{\rm{z}}$ and $Sg$.

The local magnetization, the correct and anticipated one, is then expressed 
respectively as
\begin{equation*}
  g=\hat Q^{-1}\left[H_{\rm{z}}\right], \qquad   
  g^{(i)}=\hat Q^{-1}\left[H_{\rm{z}}^{(i)}\right]
  .
\end{equation*}
Whereas $g$ is initially unknown, the $g^{(i)}$ can be evaluated.
Since  $\hat Q$ is linear one has
\begin{equation*}
  H_{\rm{z}} = H_{\rm{z}}^{(i)} + \hat Q\left[g-g^{(i)}\right]  \, ,
\end{equation*}
 or
\begin{equation*}
  H_{\rm{z}} = H_{\rm{z}}^{(i)} + \hat Q\left[(1-S)(g-g^{(i)})\right]
  + \hat Q\left[S(g-g^{(i)})\right]  \, ,
\end{equation*}
and using that $(1-S)g=0$, this may be written
\begin{equation}
  \label{tmpH}
  H_{\rm{z}}  = H_{\rm{z}}^{(i)} - \hat Q\left[(1-S)g^{(i)}\right]+ \hat Q\left[S(g-g^{(i)})\right] 
  .
\end{equation}
As a first iterative step, we neglect 
the term  in (\ref{tmpH}) containing the deviation $g-g^{(i)}$, and 
label the new approximation by $H_{\rm{z}}^{(i+1)}$, i.e., 
\begin{equation}
  \label{defH1}
  H_{\rm{z}}^{(i+1)}  \equiv H_{\rm{z}}^{(i)} + \Delta H_{\rm{z}}^{(i)} 
  \, ,
\end{equation}
where 
\begin{equation}
  \label{deltaH1}
  \Delta H_{\rm{z}}^{(i)} = - (1-S)\left(\hat Q\left[(1-S)g^{(i)}\right]+ C^{(i)}\right)
  \, .
\end{equation}
Here the constant $C^{(i)}$ compensates for the omitted term, and is given
the value required by flux conservation,
\begin{equation}
  \int  H_{\rm{z}}^{(i+1)} \,\rmd x \rmd y=0
  .
\end{equation}
The $H_{\rm{z}}^{(i+1)}$ is an improved approximation to
$H_{\rm{z}}$, and we repeat the whole procedure $s$ times untill
$(1-S)g^{(s)}$ becomes vanishingly small.
In this case $g^{(s)}$ gives the correct magnetization distribution, 
and we have successfully inverted the Biot-Savart law.

\subsection{Test case: Array of superconducting strips}

\begin{figure}[t]
  \centering  
  \includegraphics[width=13cm]{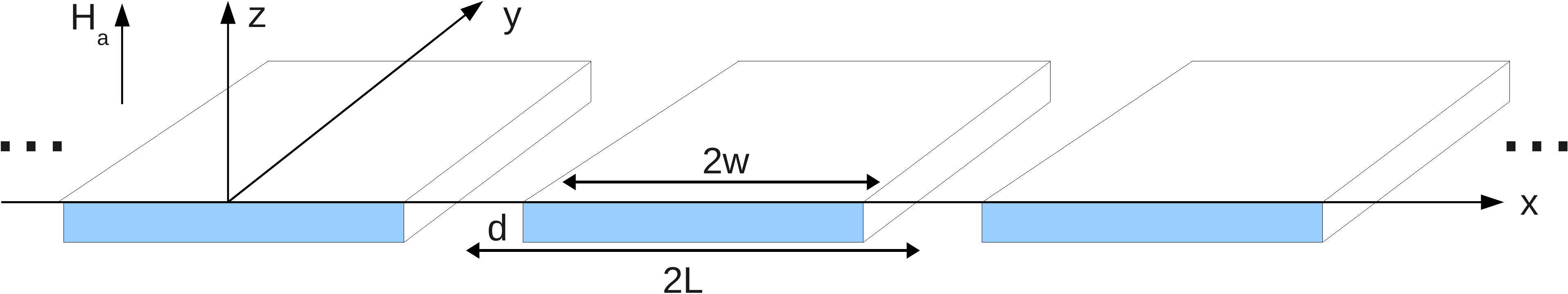} \\
  \caption{
    \label{fig:x-array}
    An x-array of strips with thickness $d$, width $2w$, 
    and center-to-center distance $2L$ placed in a transverse applied magnetic field.
  }
\end{figure}

In order to illustrate the iteration scheme, the algorithm will first be
applied to a reference case with known analytical solution.  We
consider a periodic arrangement of infinitely long superconducting
strips in the Bean critical-state, where an exact solution was
obtained by Mawatari \cite{mawatari96}. The configuration is seen in
figure~\ref{fig:x-array}, where three strips in an infinite array are
shown.  Each strip has width $2w$, thickness $d$ and center-to-center
distance $2L$. Due to the magnetic field applied in the $z$-direction,
the magnetic flux penetrates from both sides of the strips.
For the strip centered at $x=0$, the flux front position is at $|x|=a$,
and the magnetic flux distribution is
given by
\begin{equation}
  \label{H-bean}
  H_{\rm{z}}(x) = 
  H_{\rm{c}}
  \left\{
  \begin{array}{ll}
    0,                        &|x| < a, \\
    \artanh(1/|\varphi(x)|),  &a < |x| < w,\\
    \artanh|\varphi(x)|,      &w < |x| < L,
  \end{array}
  \right.
\end{equation}
where  $H_{\rm{c}}=J_{\rm{c}}/\pi$.
The corresponding sheet current is 
\begin{equation}
  \label{j-bean}
  J_{\rm{y}}(x) = J_{\rm{c}}
  \left\{
  \begin{array}{ll}
    -\frac{2}{\pi}\arctan\varphi(x),  &|x|<a, \\
    -\sgn(x),                         &a<|x|<w,
  \end{array}
  \right.
\end{equation}
where the function $\varphi(x)$ is
\begin{equation}
  \varphi(x) = \frac{\tan(\frac{\pi x}{2L})}{\tan(\frac{\pi w}{2L})}
  \sqrt{\frac{\tan^2(\frac{\pi w}{2L})- \tan^2(\frac{\pi a}{2L})}
    {|\tan^2(\frac{\pi a}{2L}) - \tan^2(\frac{\pi x}{2L})|}}
  .
\end{equation}
The width of the fluxfree area, $2a$,  shrinks with the increasing 
applied field according to
\begin{equation}
  \sin\left(\frac{\pi a}{2L}\right)
  =\frac{\sin(\frac{\pi w}{2L})}{\cosh(\frac{H_{\rm{a}}}{H_{\rm{c}}})}
  .
\end{equation}

Let us now assume that the magnetic field distribution, equation~(\ref{H-bean}), is 
known over the area of the strip, $ |x| < w $, and based only on that, 
set out to determine both the sheet current,  $ J$, and local magnetization, $g$.
We use (\ref{defH1}) and iterate
over an area $2L\times 2L$, discretized on a $256\times 256$ equidistant grid.
The calculations were performed using
$L=1.5$ and $H_{\rm{a}}=0.5$ in units where $J_{\rm{c}}=w=1$. 
As initial guess we set $H_{\rm{z}}^{(0)}(x)=$const. in the area 
between the strips.

\begin{figure}[t]
  \centering  
  \includegraphics[width=4.15cm]{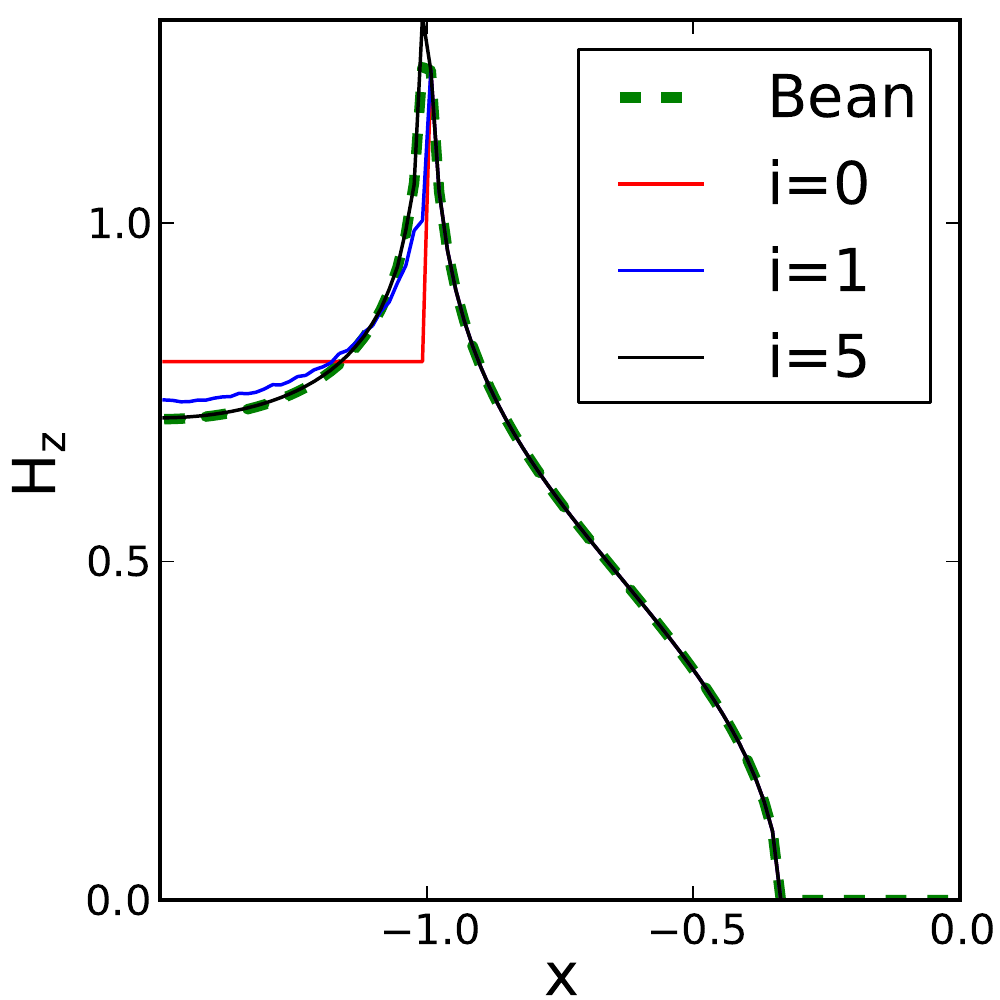} \
  \includegraphics[width=4.15cm]{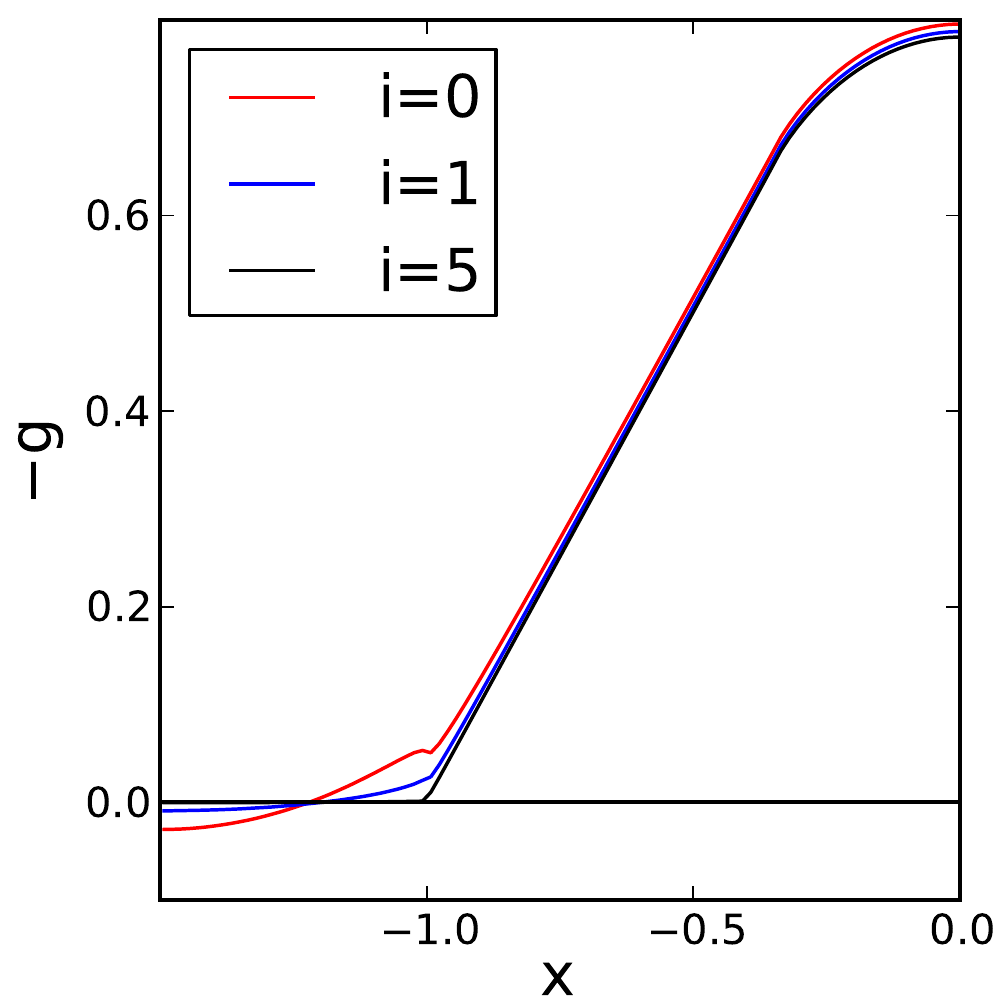} \
  \includegraphics[width=4.15cm]{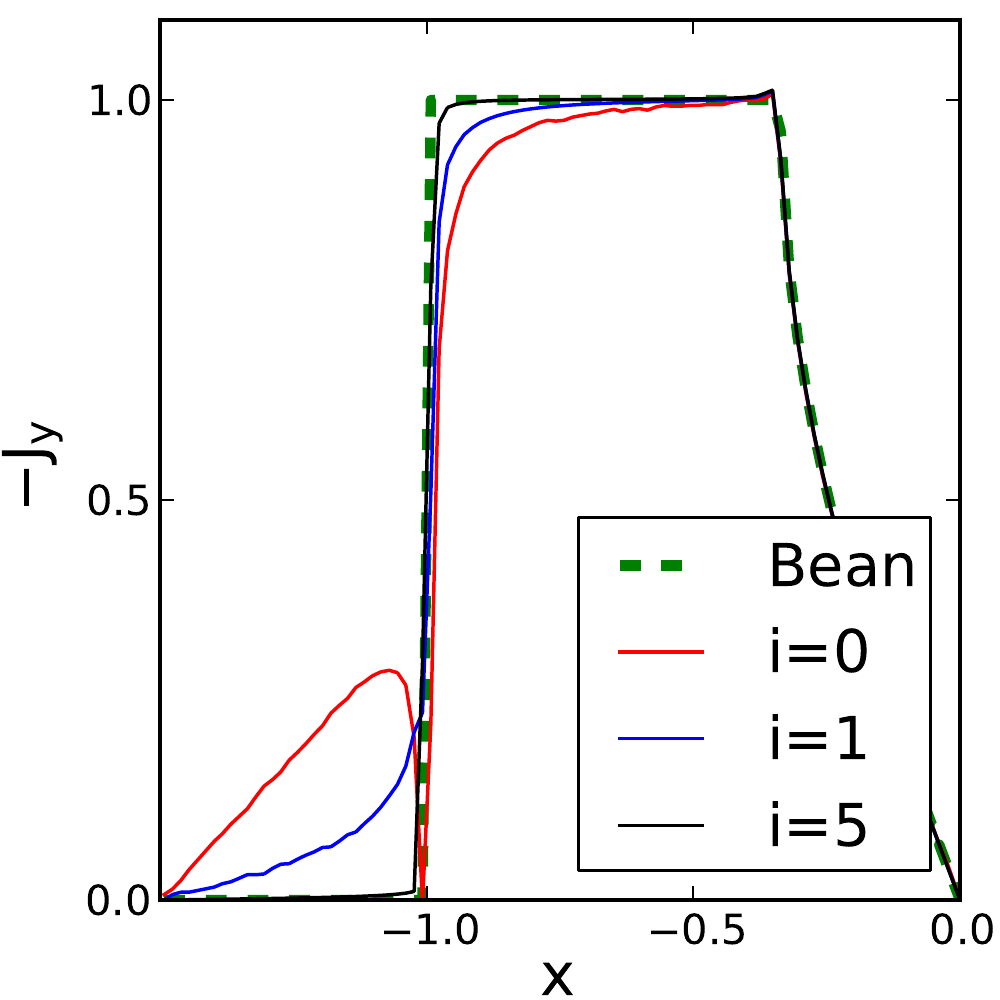} \
  \caption{
    \label{fig:check-boundary-strip}
    The iterative scheme for inversion of the Biot-Savart law 
    compared to Mawatari's analytical result.
    For increasing number of iterations $i=0$, 1 and 5, the 
    magnetic field $H_{\rm{z}}$, local magnetization $g$, and sheet current $J_{\rm{y}}$
    are closer to the analytical results. At $i=5$ 
    they are almost identical.
  }
\end{figure}

The results obtained after 0, 1 and 5 iterations, are presented in
figure~\ref{fig:check-boundary-strip}.  In spite of a poor initial guess
for the outside field, already after one iteration, the result is very
much improved. The largest deviation is that significant currents flow
in the region between the strips. However, after 5 iterations this
unphysical feature is negligible, and the numerical and exact
solutions are practically the same.  Thus, we
conclude that our iterative inversion scheme 
is very rapidly converging towards the exact
solution in this non-trivial test case.

\section{Flux dynamics}  
\label{sec:creep}

\subsection{Superconducting films }
\label{subsec:remanent}

Consider now the more general situation where a superconducting film
of finite size is experiencing a time-varying transverse homogeneous
applied magnetic field, $H_{\rm{a}}(t)$.  We want to calculate numerically the
electrodynamics as the field is gradually changing.  In such cases,
electrical currents will be induced in the sample, setting up their
own magnetic self-field.  The total transverse field, $H_{\rm{z}}$, has
therefore two contributions,
\begin{equation}
  H_{\rm{z}} = H_{\rm{a}} + \hat Q\left[g\right]
  ,
\end{equation}
where the last term represents the induced field (\ref{def-Q}).
Taking the time derivative and inverting this equation, one gets
\begin{equation}
  \label{dotg}
  \dot g = \hat Q^{-1}\left[\dot H_{\rm{z}}-\dot H_{\rm{a}}\right]
  .
\end{equation}

Outside the sample, $\dot H_{\rm{z}}$ is found by a boundary condition,
as described in section~\ref{subsec:scheme}. Inside the sample area, $\dot H_{\rm{z}}$ 
is found using the Faraday law, $\mu_0\dot H_{\rm{z}}=-(\nabla\times\bi E)_z$,
which combined with a material law $\bi E=\rho \bi J/d$
gives 
\begin{equation}
  \label{dotHz}
  \dot H_{\rm{z}} = \nabla\cdot (\rho\nabla g)/\mu_0d
  ,
\end{equation}
where the resistivity $\rho$ represents the material characteristics. 
The conventional material characteristic used to describe a superconductor 
in the slow dynamics, or flux creep regime, is a power law
\begin{equation}
  \label{power-law-Ej}
  \rho=\rho_0\left(\frac{H_{\rm{z}}}{H_{\rm{c2}}}\right)^m\left(\frac{J}{J_{\rm{c}}}\right)^{n-1}
  ,
\end{equation}
where $\rho_0$ is a resistivity constant, $H_{\rm{c2}}$ is the upper
critical field, and $J_{\rm{c}}$ the critical sheet current. 
The exponent $m$ is typically small, while the creep exponent $n\gg
1$. For high-$T_{\rm{c}}$ superconductors, e.g., YBa$_2$Cu$_3$O$_{\rm{x}}$, one
commonly finds $n=10-70$ \cite{zeldov90,sun91}, while  for MgB$_2$
exponents as high as $n=78$ were found at $T=25~$K \cite{thompson05}.
In conventional superconductors flux creep is not observed unless very
close to $T_{\rm{c}}$, so in simulations one may then set $n$ sufficiently large
to make creep negligible.

In this work we present simulation results for both stable flux creep
dynamics, and for the far more dramatic flux avalanche dynamics.  We
illustrate first the numerical scheme by applying it to the smooth
dynamics when the applied field is ramped from zero and up to a value
giving essentially full flux penetration, and then back again.  This
produces a remanent state which contains trapped flux, and is the
state used in Section~IV as starting point for simulations of
avalanches.  

To solve the dynamical equations numerically we convert them to dimensionless
form, assuming that $J_{\rm{c}}$ and $|\dot H_{\rm{a}}|$ are constants. Based on the 
sample half-width $w$, and the parameter
\begin{equation}
  \label{t0a}
  J_0 \equiv J_{\rm{c}}\left(\frac{dw\mu_0|\dot H_{\rm{a}}| H_{\rm{c2}}^m}{\rho_0J_{\rm{c}}^{m+1}}\right)^{\frac{1}{n+m}}
  ,
\end{equation}
we choose dimensionless quantities as
$\tilde {\hat Q}^{-1} \equiv \hat Q^{-1}/w$, 
$\tilde t \equiv t |\dot H_{\rm{a}}|/J_0$, 
$\tilde g \equiv g/(wJ_0)$, and
$\tilde H \equiv H/J_0$.
In these units the ramp rate satisfies $|\rmd\tilde H_{\rm{a}}/\rmd\tilde t| = 1$, and the only free 
parameters are the exponents $m$ and $n$. We will henceforth omit the tildes
when writing the dimensionless quantities. 
The sample of size $2\times 2$ is embedded in a $2.6\times
2.6$ square which is discretized on a $512\times 512$ equidistant grid. 
The results are obtained by solving (\ref{dotg})
with constant creep exponent $n=29$ and $m=0$.
The number of iterations in (\ref{defH1}) is $i=6$.

\begin{figure}[bt]
  \centering
  \includegraphics[width=13cm]{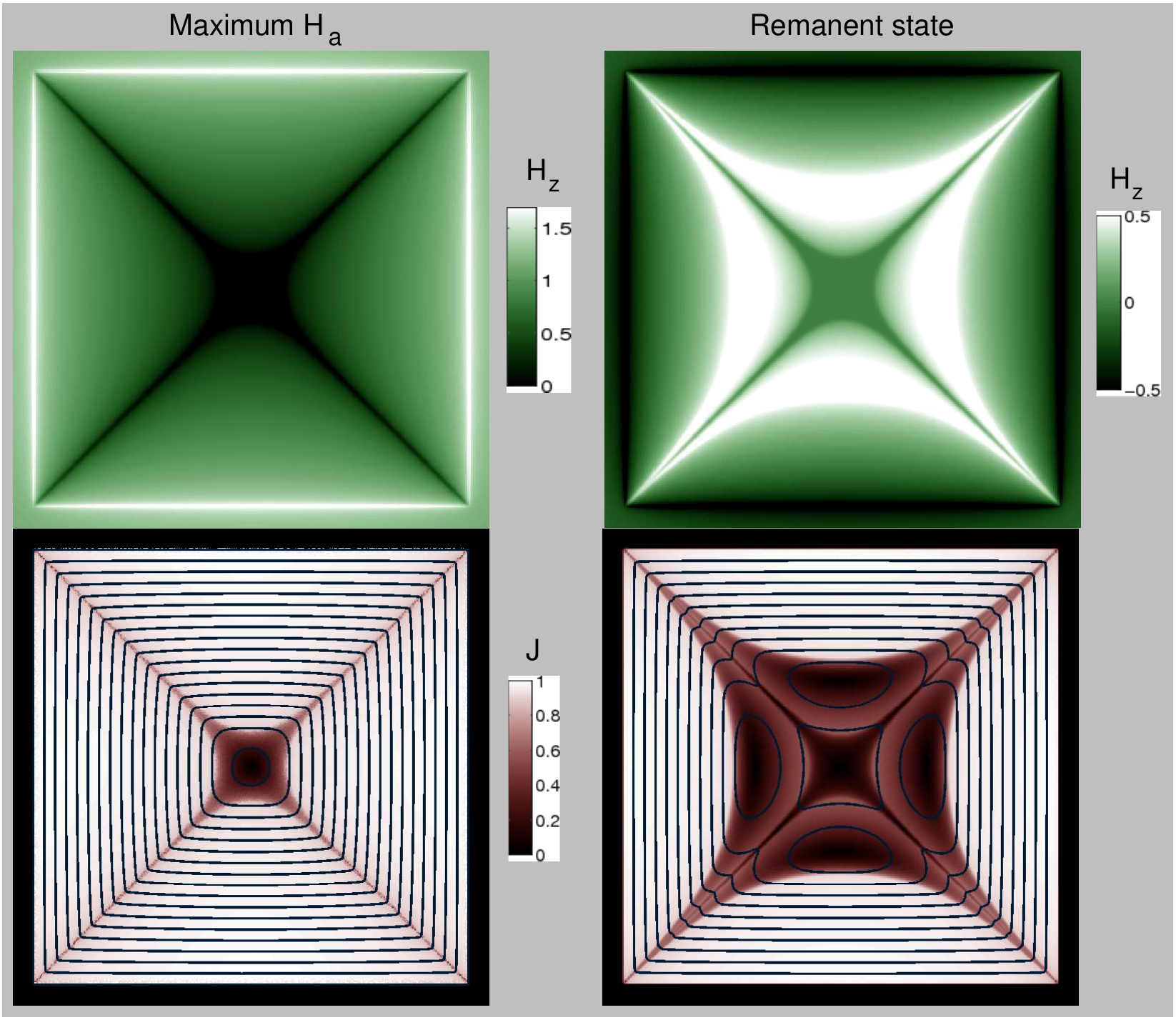} \
  \caption{
    Distributions of the magnetic field, $H_{\rm{z}}$, and electrical current magnitude, $J$,
    near full penetration (left)  and in the subsequent remanent state (right) 
    in a square superconducting film. 
    Note that the two flux density maps have different scales. 
    The current maps include also the stream line pattern of the flow.
    \label{fig:remanent}
  }
\end{figure}

Shown in figure~\ref{fig:remanent} (left) is the flux distribution and 
current stream line patterns after increasing the applied field to
$H_{\rm{a}}=1$. The field along the sample edge is much higher, reaching 
values nearly twice as large close to the mid-point of the sides.
The flux front reaches almost to the center of the sample
forming a flux density pattern often observed in MOI experiments \cite{jooss02}.
The current streamlines are in the flux-penetrated regions essentially equidistant,
and display sharp turns at the diagonals, as typical for 
a square sample in the critical state \cite{brandt95}. A slight overall convexity of the 
streamline loops is due to the finite creep exponent.

The two panels on the right show the state after the applied field was
ramped down to zero.  In this remanent state the edge field is
reversed.  As seen in the current map below the
regions of maximum current are now tongue-shaped extending from each
side of the square. Here, the nearly equidistant stream lines represent
current flow in opposite direction as compared to those in the left
panel. Only in the central part of the sample is the $\bi J$
circulating the square in the same direction as at maximum applied
field. However, the magnitude $J$ is in a large central area far below
$J_{\rm{c}}$, and the current flows in a different pattern.
The result is in good agreement with scanning Hall probe measurements
\cite{xing94}, MOI and previous numerical
simulations \cite{schuster95,baziljevich96}.

We return to this remanent state, when reporting simulations of avalanche dynamics.
Since the transient electromagnetic behavior during flux
avalanches in superconductors involves rapid localized variations in
the field taking place in normally conducting regions, we present next,
as reference, a useful exact solution to a generic dynamical problem for an
Ohmic film.

\subsection{Ohmic films}

\label{subsec:ohmic}
When a uniform magnetic field is suddenly applied transverse to a
normally conducting film, electrical currents will be induced
everywhere in the specimen. This global character of the response has
similarities to that of superconducting films. 
The case considered here, is a delta function
source-field applied instantaneously to an infinite sheet of normal
conductor, with resistivity $\rho_0$.  Let the applied field be
described by
\begin{equation}
  H_{\rm{a}} = H_0 \, \delta_2(x,y) \, \Theta(t) \, ,
\end{equation}
where $H_0$ is the field strength, $\delta_2$ is the two dimensional
delta function, and $\Theta$ is the Heaviside step function.  The
dynamical response is described by (\ref{dotg}), combined with 
Ohm's law, $\bi E=\rho_0 \bi J/d$. One then gets
\begin{equation}
 \dot g = \hat Q^{-1}\left[v_0\nabla^2g-\dot H_{\rm{a}}\right]
  \label{nonlocal-diffusion}
  ,
\end{equation}
where $\hat Q^{-1}$ is the inverse Biot-Savart operator (\ref{def-invQ}) and 
$v_0=\rho_0/(d\mu_0)$ is a constant of dimension velocity. 
Fourier transforms yield  
\begin{equation}
  -\rmi \omega \frac{k}{2} g^{[2+1]} =-v_0k^2g^{[2+1]} - H_0
  ,
\end{equation}
where $g^{[2+1]}$ is the Fourier transform of $g$ in 
two spatial dimensions plus time. Isolation of $g$ and transforming back 
to time domain for $t>0$ gives
\begin{equation}
  \label{g2}
  g^{[2]} =  -H_0\frac{2}{k}\rme^{-2v_0kt}
  .
\end{equation}
This means that the eddy currents and magnetic fields decay with 
characteristic time 
\begin{equation}
  \label{def-tau}
  \tau=l/(4\pi v_0)=\mu_0dl/(4\pi \rho_0)
  ,
\end{equation}
where $l=2\pi/k$ is the wavelength. Thus, the longest decay times are
found for the largest wavelengths.  Note that the characteristic time
for films is shorter by a factor $d/l$ compared to the slowest
decaying modes in bulk Ohmic samples \cite{ll8}.  Interestingly, 
(\ref{def-tau}) gives results in fairly close agreement with the
numerical evaluation of relaxation times after a uniform field is
abruptly applied to conducting strips and disks
\cite{brandt93-prl,brandt94-disk}.  This shows that the decay time is
only weakly sensitive to the spatial profile of the applied field
excitation as well as the shape of the Ohmic film.

Inverse Fourier transform in space of (\ref{g2}) yields the result 
\begin{equation}
  H_{\rm{z}}=H_{\rm{a}}-\frac{H_0}{\pi}\frac{v_0t}{(r^2+(2v_0t)^2)^{3/2}}
  .
  \label{Hs2}
\end{equation}
At $t\to 0^+$, the self-field is proportional to a delta function, which means that 
it shields exactly the applied field. At all times, 
(\ref{Hs2}) conserves flux, since $\int \rmd^2rH_{\rm{z}} = 0$.

The corresponding decaying sheet current is given by
\begin{equation}
  J_\varphi = \frac{H_0}{\pi}\frac{r}{(r^2+(2v_0t)^2)^{3/2}}
  .
\end{equation}
The sheet current has a maximum at 
\begin{equation}
  r_0 = \sqrt 2 v_0 t
  ,
\end{equation}
i.e. the peak moves with a constant velocity $\sqrt 2 v_0$, similar
to the eddy current front in disks after a uniform magnetic field 
is instantly applied \cite{brandt94-disk}.

\begin{figure}[t]
  \includegraphics[width=2.6cm]{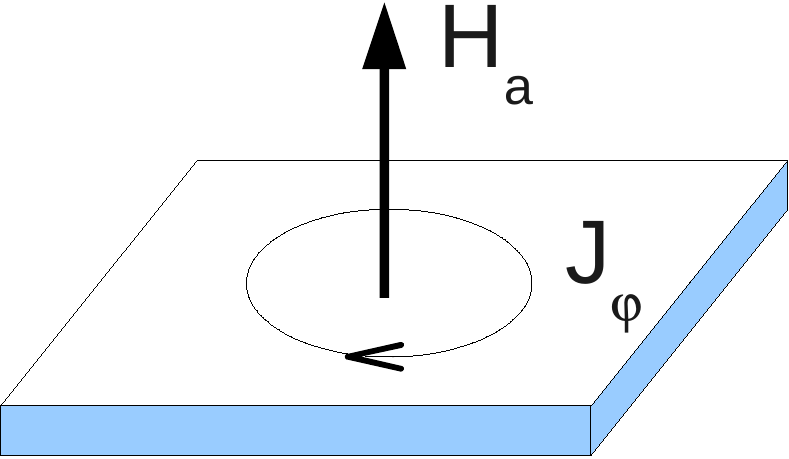} \
  \includegraphics[width=5.2cm]{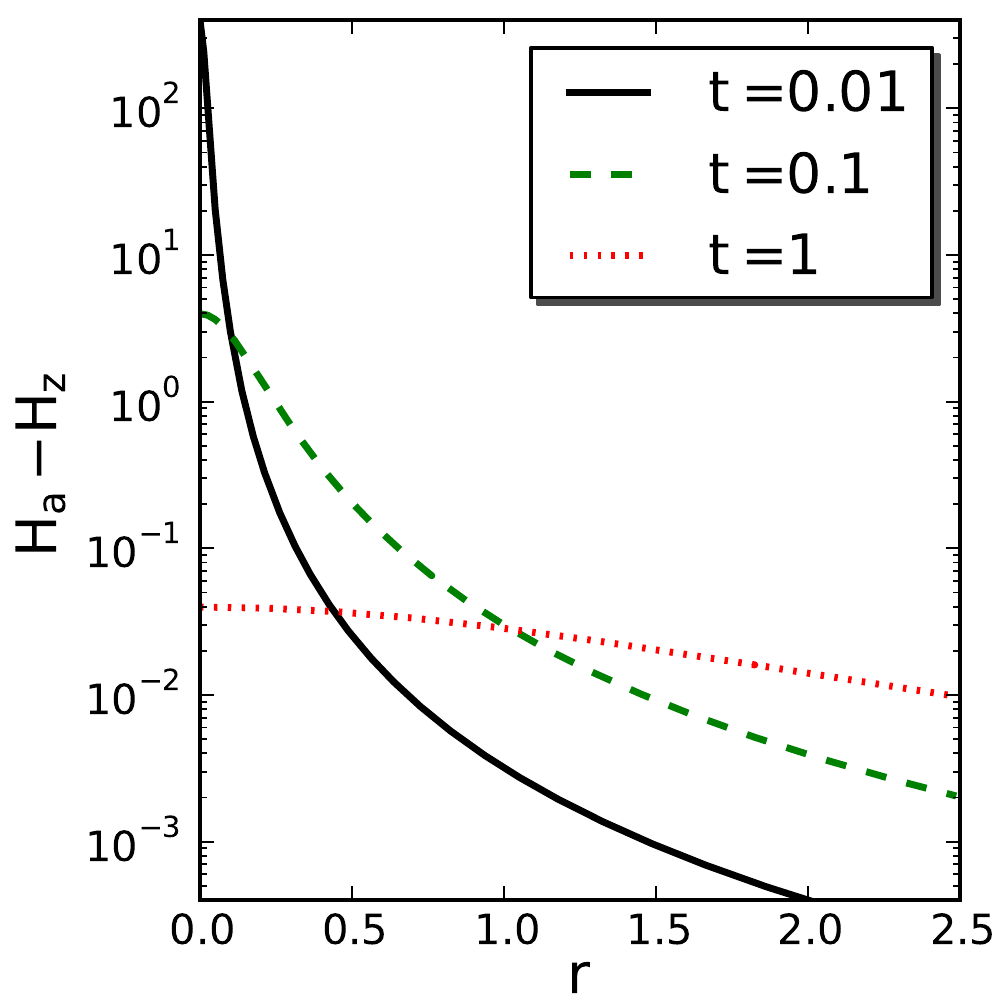} \
  \includegraphics[width=5.2cm]{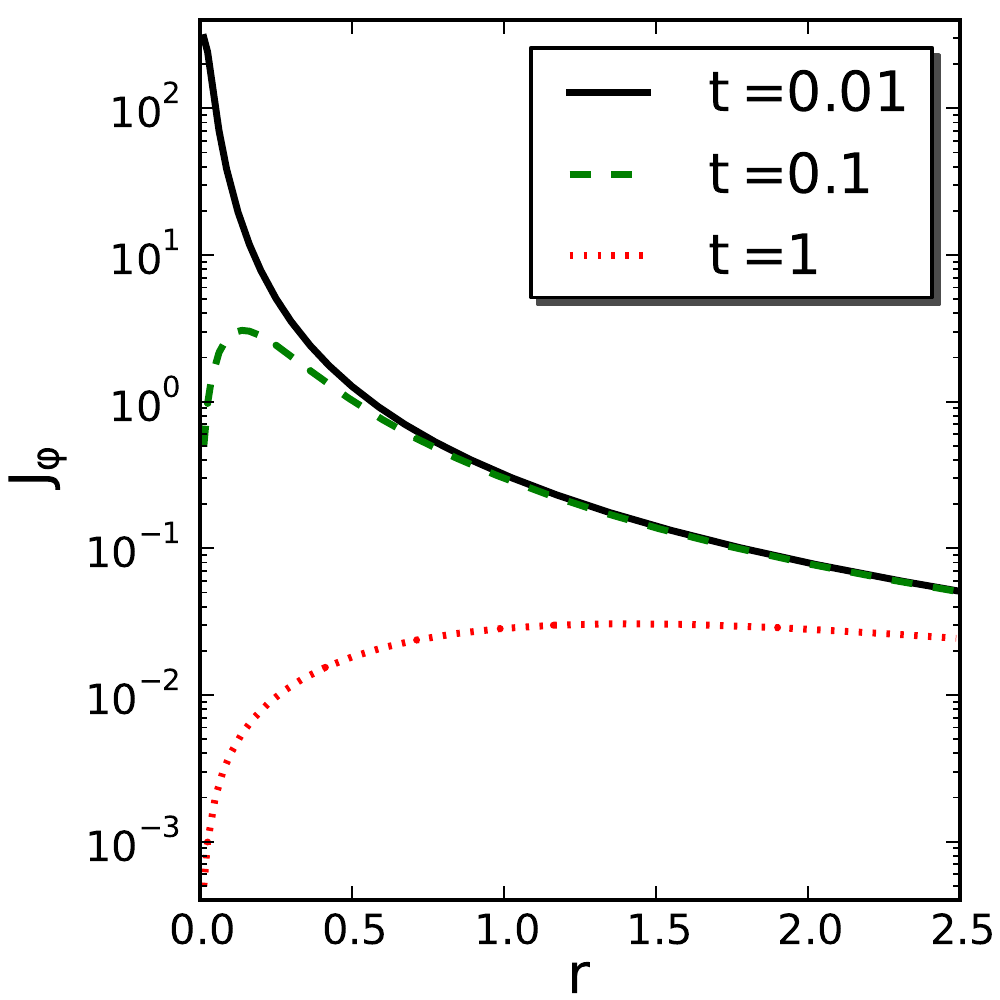} \
  \caption{
    \label{fig:hei-eddy}
    An Ohmic film exposed to a delta function source-field applied at $t=0$.
    At $t=0$ the self field $H_{\rm{z}}-H_{\rm{a}}$ shields the applied field completely, while at times 
    $0.01$, 0.1 and 1 the shielding is gradually reduced due to the decay of 
    the shielding currents $J_\varphi$.}
\end{figure}

Shown in figure~\ref{fig:hei-eddy} are the spatial profiles of the self
field, $H_{\rm{z}}-H_{\rm{a}}$, and the shielding current, $J_\varphi$, plotted at
times $t=0.01$, 0.1, and 1, in units where $H_0=v_0=1$.  In the
beginning the self field is focused near $x=0$, and with time it
decays and becomes increasingly uniform.  The result shows that
(\ref{nonlocal-diffusion}) produces a solution that is very
different from that of a diffusion process. In particular, there is no
well defined diffusion front, since both magnetic field and currents
decay algebraically as $J\sim 1/r^2$ and $H\sim 1/r^3$ at large $r$.
This contrasts the heat kernel solution of the ordinary diffusion
equation, which decays exponentially.

Although this calculation is exact only for an infinite film, 
the behavior of the shorter wavelengths should be reasonable 
approximations for finite normal domains. One such example
is the normal parts in the center of dendritic flux avalanches.
The time $\tau$ and velocity $\sqrt 2v_0$ are thus also 
expected to be characteristic
for the propagation of dendritic flux avalanches.

\section{Dendritic flux avalanches in superconductors}
\label{sec:dendrite}
Dendritic flux avalanches appear in descending as well as ascending
magnetic field. Since avalanches in descending field appear on a
highly nontrivial background, it is interesting to study their
properties. However, a numerical simulation of the full process, from
initially zero-field-cooled state, to the reentrant stability at full
penetration, and then back down, with the thermal feedback turned on,
is computationally demanding.  Therefore, we will here consider a
simpler scheme, where the remanent state is prepared with the thermal
feedback turned off, i.e., we assume that there are no avalanches, the
temperature is everywhere $T=T_0$, and the flux and current
distributions are as described in figure~\ref{fig:remanent}.  From this
background, we will explicitly trigger the dendritic flux avalanche by
a heat pulse near the edge, and consider its development.

We will in this section describe the equations governing the flow of
heat, the solution method, the units, and how to rescale the previous
result for the remanent state to these units. We will consider the
time evolution of dendritic flux avalanches nucleated at two different
locations. For comparison, we show the flux distribution of a
superconductor in descending field, mapped by the magneto-optical
imaging method.

\subsection{Preparation of the pre-avalanche state}
Due to motion of vortices there is heat dissipation in type-II
superconductors experiencing a varying external magnetic field. 
Since many of the material parameters, most notably
$J_{\rm{c}}$ and $n$, depend on temperature, the dissipation will
interfere with the electrodynamics. Thus, in order to get a complete
description of the dynamics it is necessary to model, in addition to
the electrodynamics, also the propagation of the produced heat.

Consider a superconducting film in thermal contact with a substrate of 
constant temperature $T_0$. The propagation of heat can 
then be described by the equation
\begin{equation}
  \label{dotT}
  c\dot T = \kappa \nabla^2 T - h\left(T-T_0\right)/d + JE/d
  ,
\end{equation}
where $c$ is the specific heat, $\kappa$ is the lateral thermal conductivity, $h$ is the 
coefficient of heat transfer to the substrate, and the last term represents the 
Joule heating. 
 
To transform (\ref{dotT}) into a dimensionless form, one needs to
decide on convenient scales for normalization. The most difficult scale to decide is the time scale,
since our problem is composite, with  physical processes at many different time scales. 
Under the assumption that a dendritic flux avalanches mainly propagates due to 
a domain in the normal  state invading a superconducting domain,
it is natural to chose a time scale appropriate for the decay of normal 
currents, as discussed in section~\ref{subsec:ohmic}. Thus we let 
$\tilde t=t\rho_0/dw\mu_0$, 
where $\rho_0$ is the normal resistivity of the superconductor at $T_{\rm{c}}$.
This means that the time $\tilde t\sim 1$
is characteristic for the decay of modes with size $4\pi w$.
For the other quantities, we let 
$\tilde T=T/T_{\rm{c}}$, $\tilde J = J/dj_{\rm{c0}}$, 
and $\tilde E = E/\rho_0j_{\rm{c0}}$, where  
$j_{\rm{c0}}$ is the critical current density at $T=0$.  
Since this set of units is appropriate 
for describing the fast decay of normal currents during the propagation of the dendritic 
flux avalanches, the rate of change of the applied field 
is typically very small in comparison, i.e., $\rmd\tilde H_{\rm{a}}/\rmd\tilde t \ll 1$.

The heat propagation equation then becomes
\begin{equation}
  \label{dotT2}
  \frac{\rmd \tilde T}{\rmd \tilde t} = \alpha \tilde \nabla^2 \tilde T -
  \beta (\tilde T-\tilde T_0) +\gamma\bar\gamma \tilde J\tilde E
  .
\end{equation}
Here $\alpha$ is dimensionless heat conductance, $\beta$ is
dimensionless constant for heat transfer to the substrate, and
$\gamma$ is a Joule heating parameter.
These constant parameters are defined as  
\begin{equation}
 \label{alpha-beta-gamma}
  \begin{eqalign}
    \alpha \equiv \frac{\mu_0\kappa d}{\rho_0cw},\qquad
    \beta  \equiv \frac{\mu_0wh}{\rho_0 c},\qquad
    \gamma \equiv \frac{\mu_0wdj_{\rm{c0}}^2}{T_{\rm{c}}c},
  \end{eqalign}
\end{equation}
where the material parameters at the right-hand-sides are evaluated 
at $T_{\rm{c}}$. The temperature-dependence
of $\gamma$ is taken as $\bar\gamma(T) = c(T_{\rm{c}})/c(T)$. 
In this work only the phonon contribution to $c$, 
giving $\bar\gamma = \tilde T^{-3}$ at low temperatures,
is taken into account.  
We have also assumed that the fractions 
$\kappa/c$ and $h/c$ are temperature-independent. 

Henceforth we will skip the tildes when reporting the results in 
dimensionless units.

In order to simulate 
the thermomagnetic instabilities one must specify temperature dependencies 
of $J_{\rm{c}}=dj_{\rm{c}}$ and $n$ in addition to the thermal parameters $\alpha$, $\beta$, 
and $\gamma$. We let
\begin{equation}
  J_{\rm{c}}=J_{\rm{c0}}\left(1-T\right),\qquad n = n_0/T.
\end{equation}
The resistivity is 
\begin{equation}
  \rho = 
  \left\{
  \begin{array}{ll}
    1, &
      T\geq 1 \mbox{ or }  J \geq  J_{\rm{c}}, \\
    \left(J/J_{\rm{c}}\right)^{n-1}, &
      T < 1 \mbox{ and }  J <  J_{\rm{c}}. 
  \end{array}
  \right.
  \label{rho2}
\end{equation}
Equation~(\ref{rho2}) describes a flux creep regime at $ J< J_{\rm{c}}$ and $ T<1$,
normal resistivity at $ T > 1$, and a high-resistivity flux flow regime
at $J>J_{\rm{c}}$. 
The latter implies that we have taken into account the flux flow
instability \cite{larkin75,klein85}.  This instability, which must not
be confused with the thermomagnetic instability, makes the flux flow
non-linear at high electric fields.  When the vortex velocity is
higher than a critical value, $v^*$, the resistivity jumps from the
usual flux resistive $\rho_0H/H_{\rm{c2}}$ to the much higher value
$\rho_0$.  In (\ref{rho2}), we have assumed that inside the
avalanche, the flux motion satisfies $v> v^*$, wherever $J>J_{\rm{c}}$.

The simulation of the evolution of dendritic flux avalanches 
will be based on the remanent state of figure~\ref{fig:remanent}.
It must then be transformed from units where $\dot H_{\rm{a}}=1$ to units 
with $\dot H_{\rm{a}}\ll 1$. The relevant conversion factors are 
\begin{equation}
  u \equiv J_{\rm{c}}\left(\dot{ H}_a/ J_{\rm{c}}\right)^\frac{1}{n}, \qquad
  v \equiv u/\dot{H}_a
  .
  \label{def-uv}
\end{equation}
The physical quantities will then transform as 
$g\to u g$, 
$J\to u J$, 
$H\to u H$,
$t \to v  t$, and
$E \to u E/v$.

\begin{figure*}[t]
  \centering
  \includegraphics[width=13cm]{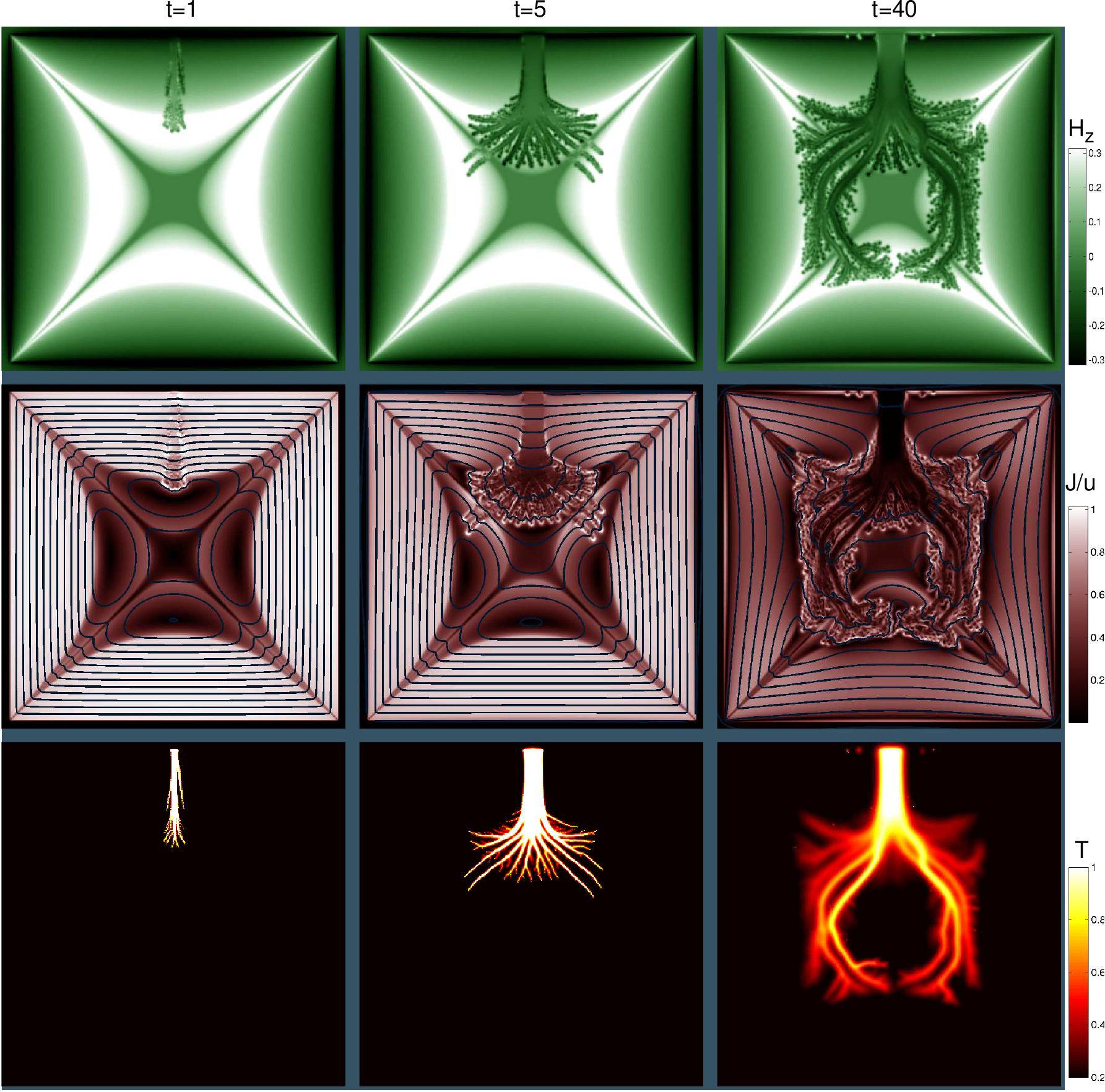}
  \caption{
    \label{fig:avalanche1}
    The development of a avalanche in the remanent state 
    showing $H_{\rm{z}}$, $J$, and $T$ at times $1$, 5 and 40
    after the nucleation.
  }
\end{figure*}

In the runs, the dimensionless parameters that characterize the thermal properties
of the sample, equation~(\ref{alpha-beta-gamma}), are selected as
$\alpha = 2\cdot 10^{-5}$, $\beta = 0.05$, $\gamma=10$.  
For the ramp rate and substrate temperature, we chose $\dot {H}_a=10^{-10}$ and $T_0=0.2$. 
This gives $J_{\rm{c}}=0.8$, and $n=100$, when $n_0=20$. The conversions factors become
$u=0.64$ and $v=8\cdot 10^9$.

The spatial disorder usually present in superconducting films
manifests itself in a non-uniform $J_{\rm{c0}}$. 
Hence, a random disorder is added to model by assigning each
grid point with $J_{\rm{c0}}\to 1 + \Delta(r-1/2)$, 
where $\Delta = 0.05$ and $r\in (0,1)$ are random numbers.

\subsection{Symmetric nucleation}
Based on the remanent state, an avalanche is nucleated centrally at one
side of the sample by assigning $T=1.5$ in a small area close to the edge.
Figure~\ref{fig:avalanche1} shows $H_{\rm{z}}$, $J$, and $T$ at times $t=1$,
$5$ and $40$ after the nucleation. 

At $t=1$ (left column of figure~\ref{fig:avalanche1})
only the critical state region is affected, and the
avalanche is mainly visible in $H_{\rm{z}}$ and $T$ as a long, thin filament
with some tendency of branching. Some of the flux is negative, which means 
that the avalanche partly consists of positive flux leaving, partly of negative flux 
entering the sample.
The heating is significant, with most of the avalanche already heated
above the critical temperature. Yet, the tip is still superconducting,
in a flux-flow state with high resistivity.  The effect on the sheet
current $J$ is less visible, although the value drops locally inside
the avalanche. As typical for the remanent state, the direction of the
$\bi J$ along the edge is such that it favors positive flux leaving and
negative flux entering the sample.

At $t=5$ (middle column) the avalanche spreads out into the inner parts of
the sample. At this stage the avalanche prefers to invade the
regions with highest flux density. The explanation of this behaviour
is in the sheet current pattern, where the branch tips are seen to
propagate transverse to the current stream lines, i.e., in the
direction of the Lorentz forces density $\bi F_L=\mu_0H_{\rm{z}}\hat z \times
\bi J$. At the same time, due to the nonlocality of the equations,
the propagating avalanche distorts the current density in a large
portion of the sample.

At $t=40$ (right column) the avalanche has essentially reached its
largest extent, and due to the efficient heat removal to substrate, the
branches are now colder.  Because of the symmetric nucleation the
avalanche is almost symmetric, but not entirely, since the state prior
to the avalanche was seeded with randomly distributed disorder.

The avalanche is large and destructive as it affects the distribution
of flux and currents in the entire sample.  Another most dramatic effect is
the strong change of the
critical state region around the edge. Before the avalanche took place, the state
was just as described by the critical state model, with constant
current density and stream lines with almost equal spacing starting
from the edge.  After the avalanche the critical state has vanished
completely, leaving a current density which is less than the half of
the original value and stream lines that are no longer parallel.  This
means that the consequences are a lot more severe for the avalanches
in the remanent state than in ascending field, where the critical
state is destroyed only in the vicinity of the avalanche
\cite{vestgarden12-sr}.

Worth noticing is also that at $t=40$, there are
small, embryonic avalanches appearing close to the edge at both sides of
the large avalanche. However, due to the above mentioned destruction
of the critical state, these are unable to develop into full
avalanches, and therefore remain small.

\begin{figure*}[t]
  \centering
  \includegraphics[width=13cm]{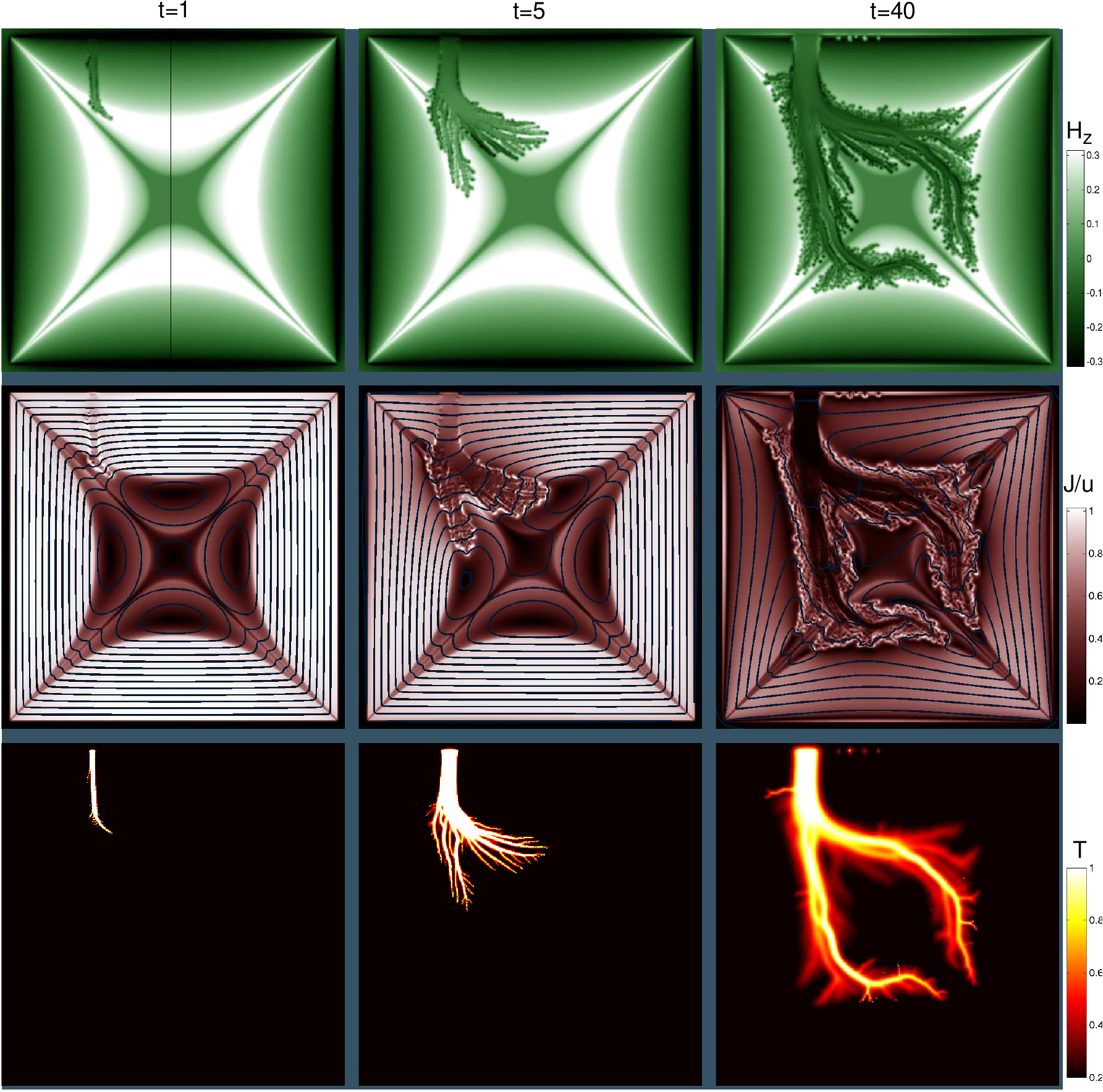}
  \caption{
    \label{fig:avalanche2}
    The development of an off-center nucleated avalanche
    showing $H_{\rm{z}}$, $J$, and $T$ at times $1$, 5 and 40
    after the nucleation. The vertical line in the upper left panel
    defines the y-axis viewed in figure \ref{fig:profileHJT}.
  }
\end{figure*}

\subsection{Off-center nucleation}
Here we investigate how 
the evolution of dendritic avalanches depends on the location
where it is initiated. We  explore this by nucleating an
avalanche away from the center of the side of the square. We use the same remanent state and
disorder configuration as for the symmetrically nucleated avalanche in
figure~\ref{fig:avalanche1}.

The results of such an asymmetrically nucleated avalanche is shown in
Figure~\ref{fig:avalanche2}, with $H_{\rm{z}}$, $J$ and $T$ obtained at $t=1$, 5, and
40.  The avalanche is nucleated close to the upper left corner, and
spreads out and fills nearly the whole inner part of the sample.
The size, shape and time evolution of the avalanche shows much
resemblance with the avalanche in
figure~\ref{fig:avalanche1}, but the symmetry of the final state is
entirely different. The final state looks like a loop, also in this
case, but it closes on the bottom right corner.

\begin{figure}[t]
  \centering
  \includegraphics[width=13cm]{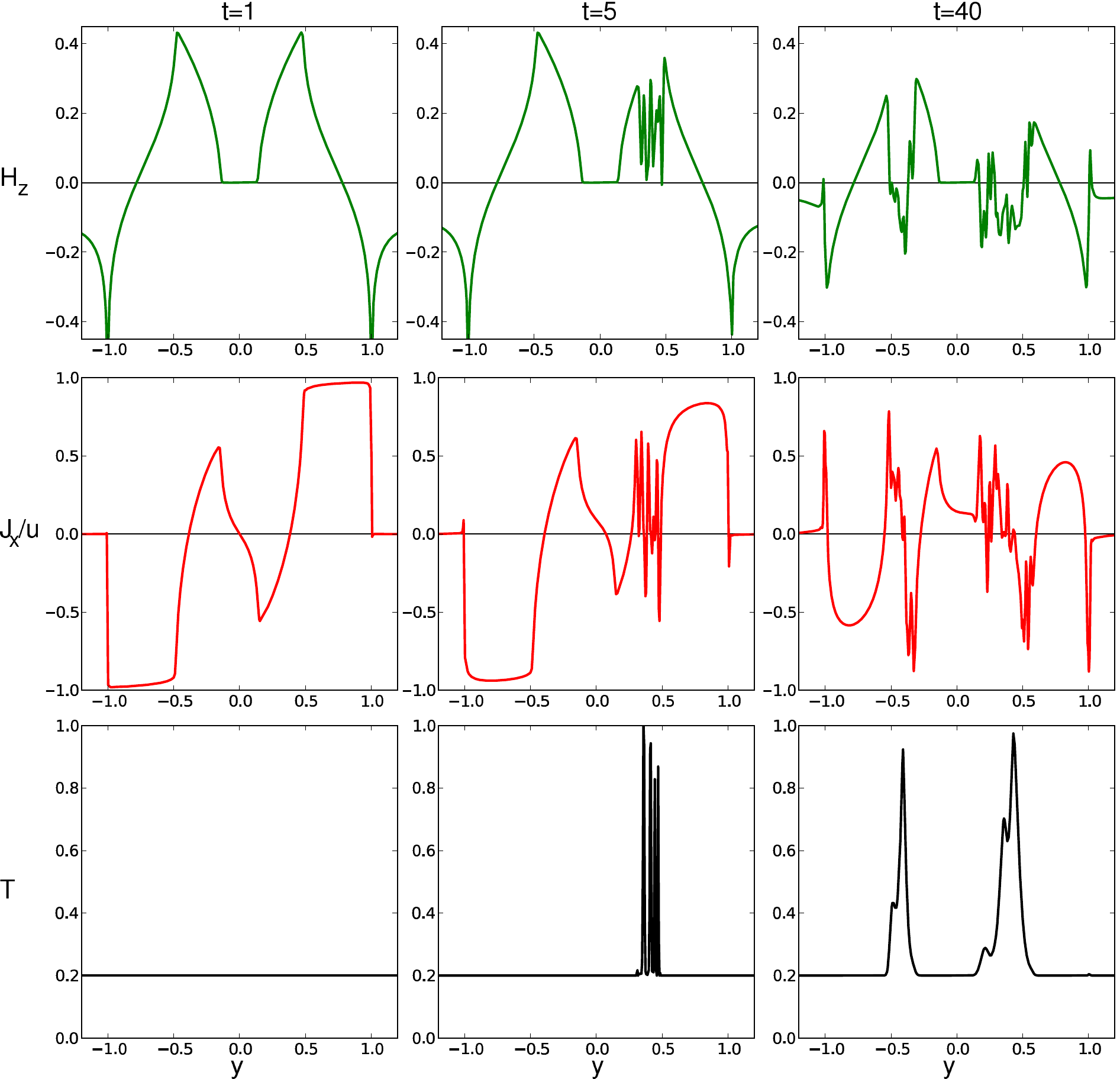} 
  \caption{
    \label{fig:profileHJT}
    The $H_{\rm{z}}$, $J$, and $T$ profiles along the y-axis at $t = 1$, 5, 40,
    for the off-center nucleated avalanche in figure~\ref{fig:avalanche2}.
  }
\end{figure}

For the off-center triggered avalanche, all the main features 
discussed for the centrally triggered avalanche are present: the enormous size,
the extensive spreading into the regions with highest flux density, the 
negative flux inside the avalanche, the destruction of the critical state, and
finally the appearance of embryonic avalanches at the edge.

Some profiles of $H_{\rm{z}}$, $J$ and $T$ along the y-axis
(vertical line through the center of the square) at times $1$, 5 and 40
are shown in figure~\ref{fig:profileHJT}. At $t=1$ all profiles are as expected 
for the remanent state in the critical state model.
In $H_{\rm{z}}(y)$ at $t=5$ one sees the finger-like 
structures penetrating the places where the flux density was highest.
The fingers consist of positive flux, while at $t=40$ there is also 
significant amounts of negative flux in the avalanche. The overall 
$|H_{\rm{z}}(y)|$ after the avalanche, both inside and outside, is much 
closer the zero than the state prior to the avalanche. 
The $J_{\rm{x}}(y)$ profiles are complex as the currents of 
the fingering structures go in opposite directions on each side 
of the fingers. More than anything, the $J_{\rm{x}}(y)$ shows that
after the avalanche event the critical state has vanished completely.
Moreover, there is no clearly preferred direction of the current.
E.g., one sees that close to the edge there is a thin layer 
with reversed current direction.
The $T$ profiles at $t=5$ shows individual hot branches 
with temperatures just below $T_{\rm{c}}$. At $t=40$ it is no
longer possible to distinguish the different branches as 
the thermal diffusion has smeared the temperature profiles.

\subsection{Magneto-optical imaging of avalanches}
\begin{figure*}[t]
  \centering
  \includegraphics[width=3.5cm]{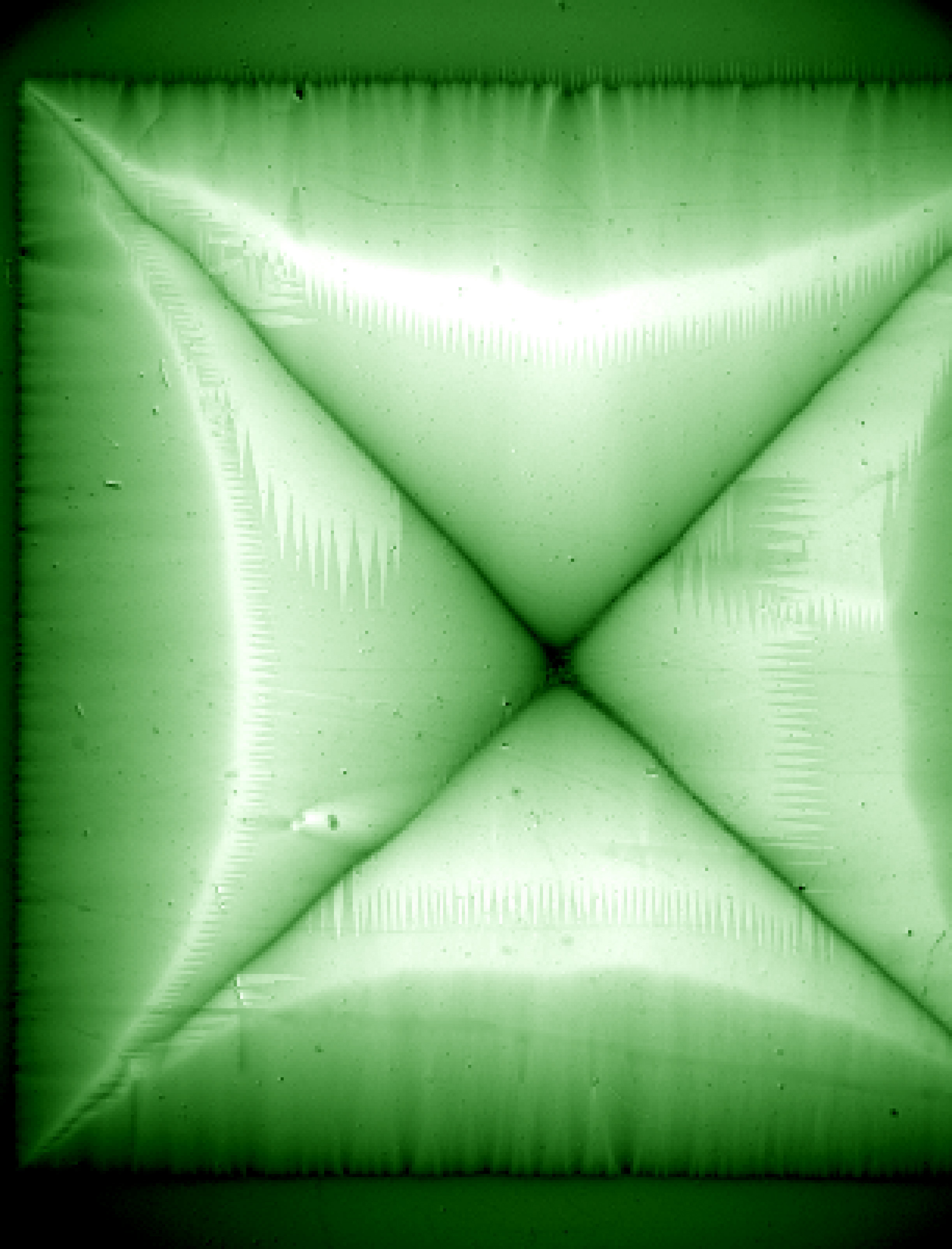} \
  \includegraphics[width=9cm]{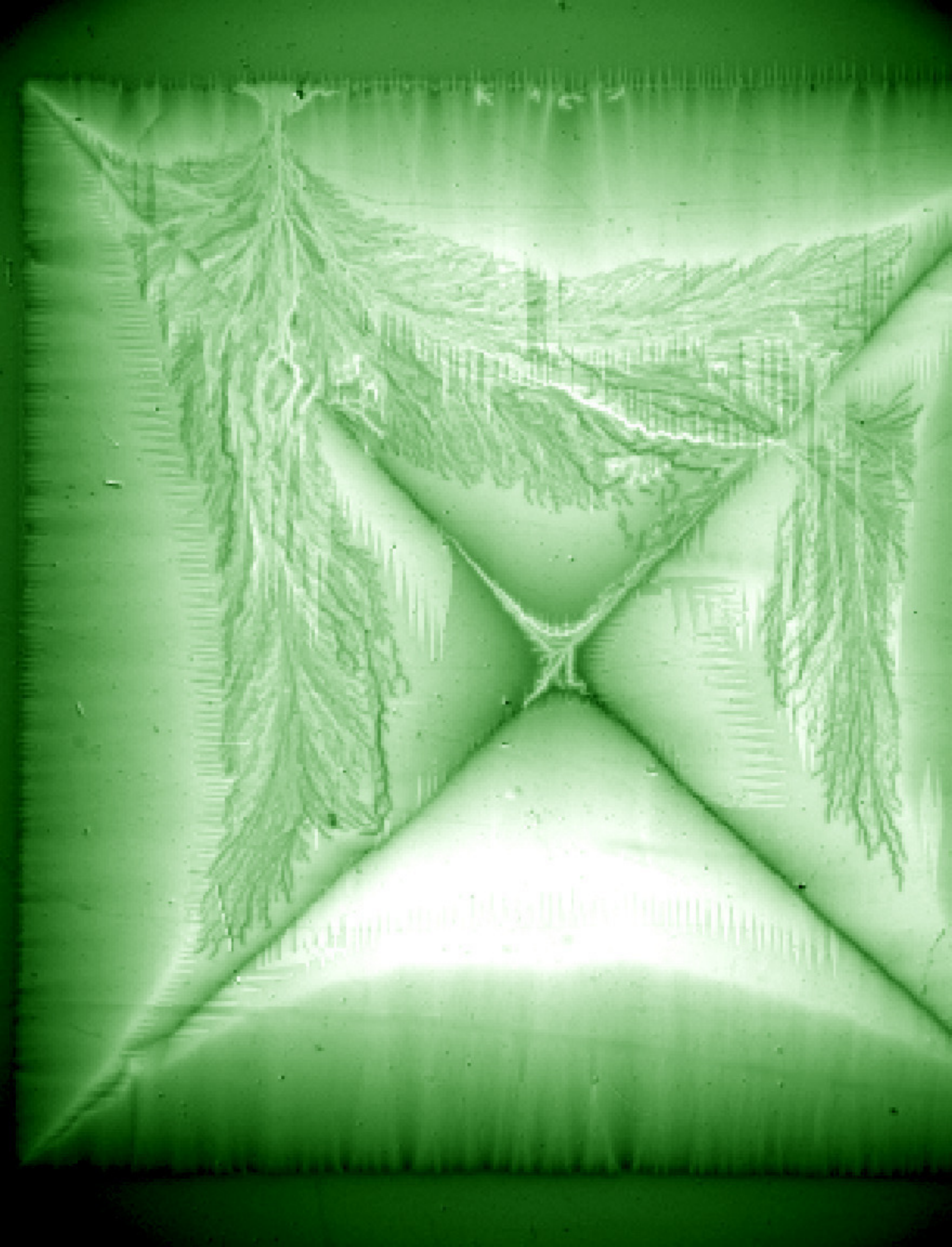} \\
  \caption{
    \label{fig:MO1}
    The state before (left) and after (right) a large dendritic flux avalanche 
    in a NbN film, 
    in descending applied field, mapped with magneto-optical imaging technique. 
  }
\end{figure*}

\label{sec:experiment}
In order to validate the correctness of the numerical solution of the
dendritic flux avalanches in the remanent state, magneto-optical
imaging experiments were performed.  The sample was a 180~nm thick NbN
superconducting film shaped as a square of sides 5.35~mm. Placed on top of the sample 
was an in-plane magnetization ferrite garnet film used as 
Faraday-rotation sensor \cite{helseth02}.
Since the Faraday rotation increases monotonously with
the perpendicular component of magnetic field, one can by polarized light
microscopy create a map of the magnetic field distribution above
the film \cite{johansen96}.

The sample was initially zero-field-cooled to 4~K and magnetic field
was applied perpendicular to the film. During the ascending field ramp,
there were many avalanches, but due to the reentrant stability in high
fields, the full penetration state at $17~$mT was critical state 
like \cite{yurchenko07}.  In
descending field, the flux dynamics was for a long time smooth and at
10.5~mT the flux distribution was as shown in the small image in
figure~\ref{fig:MO1}. Then, suddenly, a large avalanche stroke,
and in a short time, 
it entered a large portion of the sample. This large
avalanche, seen in the main image of figure~\ref{fig:MO1}, is 
typical for avalanches in descending field, near the 
avalanche threshold temperature \cite{wimbush04}.

Although the avalanche did not strike exactly in the remanent state
it is close enough to be used in a qualitative comparison with the
simulation.  First we note that the avalanche has a clear similarity to
the simulated flux avalanche in figure~\ref{fig:avalanche2}, as it
avoids the critical state region close to the edges and instead it
invades the region with highest magnetic flux density.  The size and
extent of the avalanche is also similar. The majority of branches are
dark meaning that the flux density is low.  Some of the branches are
white. In this case it is not clear if this means negative flux, as was
reported in the simulations, or positive flux, since the image 
only shows the absolute value of $H_{\rm{z}}$. One more detail worth
noticing, is the appearance of embryonic avalanches at the edge of 
the sample, just as predicted by the numerical simulations.

The main discrepancy between the flux distribution of the simulation
and the magneto-optical experiment is the width of the branches.  In
the experiment they are much more narrow than in the simulation.  This
is an indication that the NbN film has lower value of the
effective heat diffusion parameter $\alpha$, given in (\ref{alpha-beta-gamma}), 
than what was used in the simulation. 
However, due to the limited spatial resolution
one cannot run the simulations with smaller values of $\alpha$ without
at the same time increasing the number of grid points. 

\section{Film with antidots}
\label{sec:antidot}
The formalism described in section~\ref{sec:model} for modelling 
the dynamics of thin-film superconductors in transverse field, 
is valid only for simply connected samples. We will now extend
the formalism to multiply connected samples. This gives us the opportunity 
to study also the flux dynamics of superconducting films with antidots (non-conducting holes).
This is of interest since, 
due to the nonlocal electrodynamics, the presence of antidots may strongly influence the 
distribution of flux and current in the films. For example, it has been reported 
that patterning with regular arrays of antidots 
makes the magnetic flux penetration anisotropic \cite{pannetier03,tamegai10}.
Currently, there are only few numerical simulation works that has considered 
the critical state flux penetration in samples patterned with antidots 
\cite{prigozhin98,crisan05,gheorghe06,vestgarden08,vestgarden12,barrett12}.
In order to improve the theoretical knowledge on the field, we 
will here consider the numerically challenging sample configuration of 
a superconducting ring patterned with a square array of disk-shaped antidots. 


\begin{figure}[t]
  \centering
  \includegraphics[width=13cm]{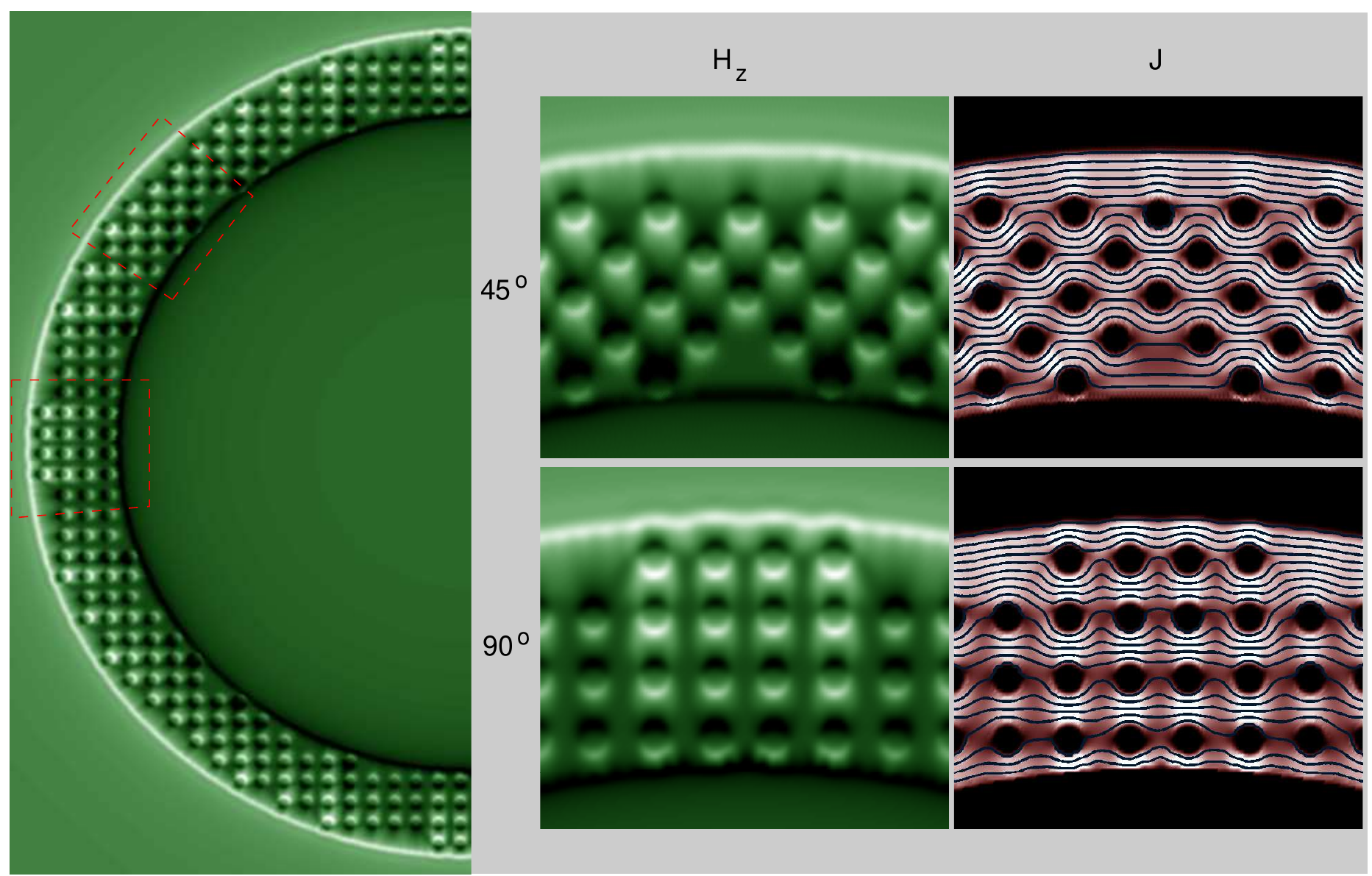} \
  \caption{
    \label{fig:ring}
    The flux and currents distribution in a ring with small round antidots 
    arranged in a rectangular pattern. The left image shows $H_{\rm{z}}$ in half of the sample,
    the central panels show close-up views of $H_{\rm{z}}$  at $45^\circ$ and $90^\circ$ orientation,
    and the right panels show the corresponding $J$.
  }
\end{figure}

Let us first consider the boundary conditions.  When the film contains
holes of any shape and number, their presence can be implemented by an
iterative scheme similar to that described
in section \ref{subsec:scheme}. 
For each hole, labeled $\alpha = 1 \ldots N$, 
we then define the hole projection 
\begin{equation}
  \label{defhm}
  h_\alpha(x,y) = 
  \left\{
  \begin{array}{ll}
    1,&  \rm{inside~hole~}  \alpha, \\
    0,&  \rm{outside~hole~}  \alpha .
    \end{array}
    \right.
\end{equation}
Equation~(\ref{defH1}) now becomes
\begin{equation}
  H_{\rm{z}}^{(i+1)}  =  H_{\rm{z}}^{(i)} +   \Delta H_{\rm{z}}^{(i)}+ \sum_{\alpha} \Delta H_{\rm{z},\alpha}^{(i)} 
  \, ,
\end{equation}
where 
\begin{equation}
  \label{deltaH2}
  \Delta H^{(i)}_{\rm{z},\alpha} = -h_{\alpha}\left(\hat Q_{\alpha}\left[h_{\alpha} g^{(i)}\right] + C_{\alpha}^{(i)}\right)
  ,
\end{equation}
which allows $H_{\rm{z}}$ in each hole to be reconstructed.
The constants $C_{\alpha}^{(i)}$ are fixed by the flux conservation condition 
\begin{equation}
  \int \Delta H_{\rm{z},\alpha}^{(i+1)} \, \rmd x \rmd y=0
  .
\end{equation}

The operator $\hat Q_{\alpha}$ can be any implementation of the forward
Biot-Savart law. In general, it is beneficial to use different
implementations for large and small holes.
For large holes, the best is to let $\hat Q_m=\hat Q$, i.e., the full
Biot-Savart law, equation~(\ref{def-Q}). The drawback of 
this approach is that it runs over all grid points. 
The advantage is that the linear operator $\hat Q$ can be moved outside
the sum in (\ref{deltaH2}) when there are more than one large hole.
For small holes one can use an implementation of
$\hat Q$ which for each hole only loops over the grid points in the
hole.
For convergence of the procedure the input to the operator $\hat Q_n$
should first be shifted to minimize contributions from the edge of the hole. This will 
reduce the damaging effect of the sharp cut made by $h_m$.

Let us consider the flux penetration in a superconducting ring with
antidots patterned in a square grid. This layout allows us to
illustrate the consequences of electromagnetic non-locality and
non-trivial dynamics given the conflicting symmetries of sample and
the array of antidots. In units where the outer radius is $R=1$ and
$\dot H_{\rm{a}}=1$ (same as the remanent state of section \ref{subsec:remanent}), the inner
radius is 0.8, and the antidots, 385 in total, have radii
$a=0.013$. The center-to-center distances of the antidots are $4a$. In
order to apply the boundary conditions, the ring is embedded in a
square of size $L_{\rm{x}}=L_{\rm{y}}=1.3$, which is discretized on a $1024\times
1024$ equidistant grid.

The left panel of Figure \ref{fig:ring} shows the flux distribution at
$H_{\rm{a}}=0.2$ when flux has fully penetrated the ring, starting to fill
the central hole with flux. The outer edge is white indicating high
flux density and the inner edge is dark indicating negative flux, as
typical for the ring geometry \cite{brandt97}. This means that the
currents flow in a clockwise direction everywhere in the ring,
contrary to a strip where currents flow in both directions.  The local
flux distribution inside the sample is much distorted due to the
presence of the antidots. The current stream lines has to bend around
the antidots and this induces large amount of flux in the
antidots. The flux is negative towards the outer edge and positive on
the other side of the antidot. For the holes closest to the inner edge
of the ring the situation is opposite.  This means that the inner edge
to large extent behaves like an outer edge subjected to a negative
applied field.

The four right panels show close-up views of $H_{\rm{z}}$ and $J$ around
$45^\circ$ and $90^\circ$ direction.  The $J$-maps show that there are
connected critical state region with $J\approx 1$ extending from the
outside to the inner edge of the ring.  The critical state connected
regions follow the symmetry of the antidot lattice and these act like
channels for easy flux penetration \cite{crisan05,vestgarden12}.
Hence, the antidot lattice makes the flux penetration anisotropic, in
good agreement with previous magneto-optical experiments on
superconducting disks patterned with antidots
\cite{pannetier03,tamegai10}.

Between the antidots there are places where $J<1$. This is
a feature that cannot be predicted by the Bean model and it shows that
it is necessary to solve the time-dependent equations to get a correct
description of the state.  These sub-critical pockets may be of technological
relevance because they imply that sample patterned with periodic
antidot arrays has better shielding properties for local magnetic
fields than unpatterned samples.

\section{Conclusions}
\label{sec:conclusion}
The macroscopic electrodynamics of thin films, either superconducting
or Ohmic, in transverse applied field can be modeled by the Maxwell
equations.  The formalism is capable of handling a wide range of
physical systems, where the material-specific properties are
introduced as an $E-J$ relation, which is linear for Ohmic conductors,
nonlinear for superconductors.  A challanging point in the formalism is to
calculate the currents for a known distribution of the magnetic
field.  We solve this problem by a hybrid real space - Fourier
space iterative scheme, which is both computationally efficient and is 
able to handle also samples with non-symmetric boundary.

When magnetic field is increased to reach full flux penetration and
then decreased to zero, superconductors with strong flux pinning
experience that a large amount of remanent flux is trapped inside the
specimen.  Both the distributions of current and magnetic field in
this remanent state are highly nontrivial, as we showed by a numerical
simulation on a film with square shape. In order to consider 
how dendritic flux avalanches evolve on the
background of the remanent state, we developed the formalism for rescaling 
solutions and for calculating the flow of heat.  The dendritic flux avalanche in the remanent
state was found to develop as an irregular branching structure that
enters the inner parts of the sample.  The avalanche consisted partly
of positive flux leaving the sample, partly of negative flux entering.
It was found to be more destructive than avalanches in ascending field
since, after the avalanche, the critical state had vanished completely
from the entire film.  The spatial extent of the avalanches was
sensitive to the nucleation position, but the size and overall
consequences were not.  A magneto-optical imaging experiment on
dendritic flux avalanche in descending field in a NbN film showed
similar looking gigantic flux avalanches and supported the findings of
the simulations.

Very few problems related to the nonlocal electrodynamics of thin films are 
analytically solvable. An exception is the response of an infinite Ohmic 
film to a delta function source field, turned on instantly. We calculated this 
solution, both in Fourier and real space, and the solution gave much insight 
into the behaviour of Ohmic films or Ohmic domains in transverse field.
We found that, in Fourier space, the mode of wavelength $l$ decays with characteristic time
$\mu_0dl/(4\pi \rho_0)$. 
The solution in real space showed that there was no well-defined 
front of propagation, 
since both current and magnetic self-field decreased algebraically 
far from the source. Yet, the current had a maximum moving away from 
the source with constant velocity $\sqrt 2\rho_0/(d\mu_0)$.

Finally, we generalized the numerical simulation formalism from 
simply connected to multiply connected geometry, i.e., we allow the samples to 
contain non-conducting holes. Thereupon, we consider the magnetic flux 
penetration in a superconducting ring with antidots (small holes)
distributed  in a square array. The magnetic flux distribution was locally 
much perturbed by the antidots, and also the large scale flux distribution 
was modified, as it became anisotropic when the magnetic flux was guided along the directions
of the antidot array. Between the antidots there were localized regions with low 
flux traffic and $J<J_{\rm{c}}$. This is contrary 
to the situation in simply connected samples, where the regions with 
$J<J_{\rm{c}}$ usually form large connected domains. The current distribution
inside the ring patterned with antidots was thus highly nontrivial,
even in the critical state.

In summary, we have shown that a wide range of apparently different phenomena related 
to the electrodynamics of superconducting and Ohmic films in transverse field can be 
described by one formalism based on the Maxwell equations and material-specific $E-J$ 
relations.

\ack
This work was supported financially by the Research Council of Norway.

\section*{Appendix}
\label{sec:appendix}
Here we give some hints and tricks for the implementation of the numerical scheme.

Both the thermal diffusion equation and the electrodynamics are
discretized on an equidistant rectangle of size $2L_{\rm{x}}\times 2L_{\rm{y}}$ 
with points
$x_i=(2i-N_{\rm{x}}+1)L_{\rm{x}}/N_{\rm{x}}$ and 
$y_j=(2j-N_{\rm{y}}+1)L_{\rm{y}}/N_{\rm{y}}$
where $i=0...N_{\rm{x}}-1$ and $j=0...N_{\rm{y}}-1$.
The discrete wave vectors that
are used in the fast Fourier transforms are 
$k_{\rm{x},p}=p\pi/L_{\rm{x}}$ and 
$k_{\rm{y},p}=q\pi/L_{\rm{y}}$,
for $p=-N_{\rm{x}}/2...N_{\rm{x}}/2-1$ and $q=-N_{\rm{y}}/2...N_{\rm{y}}/2-1$ and 
$k_{pq}=\sqrt{(k_{\rm{x},p})^2+(k_{\rm{y},q})^2}$.
Before the wave vectors can be used in the direct products of
equations (\ref{def-Q}) and (\ref{def-invQ}) the Brillouin zones must be
rearranged to ensure that the product satisfies the symmetry
conditions $\chi_{m,n}=\chi_{N_{\rm{x}}-m,n}^*$ and
$\chi_{m,n}=\chi_{m,N_{\rm{y}}-n}^*$, which are valid for the Fourier
component $\chi$ of any real function.  This means that the Fourier
transform can be optimized by keeping only half the Fourier components
and acquiring the rest by symmetry considerations.

The diffusion equation tends to be numerically unstable 
when solved forward in time. Thus we solve it by a Forward-backward average scheme.
Let $\dot T(t)\to (T^{(n+1)}-T^{(n)})/\Delta t_n$ and $T(t) \to (T^{(n+1)}+T^{(n)})/2$,
where $T^{(n)}=\mathcal F\left[T(t_n)\right]$, $t_n$ is the discrete time,
and  $\Delta t_n = t_{n+1}-t_n$.
Inserting this into (\ref{dotT2}) and isolating $T^{(n+1)}$ gives
\begin{equation}
  \label{discreteT}
\fl
  T^{(n+1)} =
  \frac{1-\left(\alpha k^2
        +\beta \right)\Delta t_n/2}
  {1+\left(\alpha k^2+\beta\right)\Delta t_n/2} T^{(n)}  
  +\frac{\mathcal F\left\{\gamma\bar\gamma JE + \beta T_0\right\}
   }
  {1+\left(\alpha k^2+\beta \right)\Delta t_n/2}  
    \Delta t_n
   .
\end{equation}
This equation is finite in both limits $k^2\to 0$ and $k^2\to \infty$,
contrary to a forward-in-time integration scheme,  $ T(t) \to  T^{(n)}$,
which diverge as $k^2\to \infty$.

Due to the nonlinearity, (\ref{dotg}) must be solved forward in
time, for example using the Runge-Kutta method.  
It is essential to use a variable time step, 
$\Delta t \propto 1/E_{\max}$, where $E_{\max}$ is the global
maximum value of the electrical field \cite{brandt95}.  The non-locality of the
equations implies that there is only one global time step, selected 
by considering the most pronounced flux traffic.

A delicate point in the numerical simulation scheme is the  
execution of the spatial derivatives in (\ref{dotHz}). 
This work applies a finite difference randomly alternating 
between $(f(x_{i+1})-f(x_i))/(x_{i+1}-x_i)$ and $(f(x_{i})-f(x_{i-1}))/(x_{i}-x_{i-1})$.
This produces stable results and gives no systematic error.
 
The inverse Biot-Savart law, equation~(\ref{def-invQ}), is divergent at
$k_{00}=0$. This reflects the fact that $k_{00}$
describes a spatially constant mode, while $g$ is only defined up to
a constant by $\bi J=\nabla\times \hat zg$. However,
the constant is fixed by the requirement that the magnetic moment of 
the vacuum is zero. This boundary condition is most easily implemented 
in real space. Hence, we assign $k_{00}=1$ to avoid the singularity,
then shift the output to satisfy $\int \rmd^2r (1-S)g = 0$. 

One threat to the convergence of the iterative scheme of equation~(\ref{defH1})
is the discontinuity of the projection $1-S$. A counter-measure 
is to smoothen the output of $\hat Q$ by a multiplication 
by a Gaussian in Fourier space,
\begin{equation}
   \hat{Q}\left[g(\bi r)\right] 
  = \mathcal{F}^{-1}
  \left[\frac{k}{2}\mathcal{F}\left[g(\bi r)\right]
    \exp\left(-\frac{1}{2}\sigma^2 k^2\right)
  \right]
  .
\end{equation}
In real space, this implies a convolution with a Gaussian
\begin{equation}
  \Phi_\sigma(\bi r) = 
  \frac{1}{2\pi \sigma^2}
  \exp\left(-\frac{1}{2}\left(\frac{r}{\sigma}\right)^2\right)
  ,
\end{equation}
i.e., the result is an interpolation with a neighborhood of size $\sigma$.
It is reasonable to let $\sigma$ be a small number
of order the grid size $\sigma \sim 2a/N_{\rm{x}}$. Note that 
$\lim_{\sigma\to 0}\Phi_\sigma=\delta(x)\delta(y)$.
The same Gaussian smoothing should also be applied to $\hat Q^{-1}$.

\bibliographystyle{unsrt}

\bibliography{superconductor}

\end{document}